\documentclass[useAMS,usenatbib]{mn2e}
\usepackage{graphicx} 
\usepackage{epsfig}
\usepackage{deluxetable}
\usepackage{rotating}
\usepackage{times}
\usepackage{amssymb,amsmath}

\title[Unveiling the nature of the $\gamma$-ray emitting AGN PKS 0521$-$36]
{Unveiling the nature of the $\gamma$-ray emitting AGN PKS 0521$-$36}

\author[F. D'Ammando, M. Orienti, F. Tavecchio, et al.]{F. D'Ammando$^{1,2}$\thanks{E-mail: dammando@ira.inaf.it}, M. Orienti$^{2}$, F. Tavecchio$^{3}$, G. Ghisellini$^{3}$, E. Torresi$^{4}$, M. Giroletti$^{2}$, \newauthor C. M. Raiteri$^{5}$, P. Grandi$^{4}$, M. Aller$^{6}$, H. Aller$^{6}$, M. A. Gurwell$^{7}$, G. Malaguti$^{4}$, E. Pian$^{8}$, \newauthor G. Tosti$^{9,10}$\\
$^{1}$Dip. Di Fisica e Astronomia, Universit\`a degli Studi di Bologna, Via Ranzani 1, I-40127, Bologna, Italy\\
$^{2}$INAF - Istituto di Radioastronomia, Via Gobetti 101, I-40129 Bologna, Italy\\
$^{3}$INAF - Osservatorio Astronomico di Brera, via E. Bianchi 46, I-23807 Merate, Italy \\
$^{4}$INAF - Istituto di Astrofisica Spaziale e Fisica Cosmica, Via Gobetti 101, I-40129 Bologna, Italy\\
$^{5}$INAF - Osservatorio Astrofisico di Torino, Via Osservatorio 20, I-10025 Pino Torinese (TO), Italy\\
$^{6}$University of Michigan, Ann Arbor, MI, USA \\
$^{7}$Harvard-Smithsonian Center for Astrophysics, Cambridge, MA 02138, USA \\
$^{8}$Scuola Normale Superiore Pisa, Piazza dei Cavalieri 7, I-56126 Pisa, Italy \\
$^{9}$Dip. di Fisica, Universit\`a degli Studi di Perugia, Via A. Pascoli, I-06123 Perugia, Italy \\
$^{10}$INFN - Sezione di Perugia, Via A. Pascoli, I-06123 Perugia, Italy \\
}

\begin{document}

\date{Accepted. Received; in original form}

\maketitle

\begin{abstract}
PKS 0521$-$36 is an Active Galactic Nucleus (AGN) with uncertain classification. We investigate the properties of this source from radio to
$\gamma$ rays. The broad emission lines in the optical and UV bands and steep
radio spectrum indicate a possible classification as an intermediate object
between broad-line radio galaxies (BLRG) and steep spectrum radio quasars
(SSRQ). On pc-scales PKS 0521$-$36 shows a knotty structure similar to
misaligned AGN. The core dominance and the $\gamma$-ray properties are similar
to those estimated for other SSRQ and BLRG detected in $\gamma$ rays, suggesting an intermediate viewing angle with respect to the observer. In this context the flaring activity detected from this source by {\em Fermi}-LAT between 2010 June and 2012 February is very intriguing. We discuss the $\gamma$-ray emission of this source in the framework of the structured jet scenario, comparing the spectral energy distribution (SED) of the flaring state in 2010 June with that of a low state. We present three alternative models corresponding to three different choices of the viewing angles $\theta _{\rm v}$ = 6$^{\circ}$, 15$^{\circ}$, and 20$^{\circ}$. We obtain a good fit for the the first two cases, but the SED obtained with $\theta _{\rm v}$ = 15$^{\circ}$ if observed at a small angle does not resemble that of a typical blazar since the synchrotron emission should dominate by a large factor ($\sim$ 100) the inverse Compton component. This suggests that a viewing angle between 6$^{\circ}$ and 15$^{\circ}$ is preferred, with the rapid variability observed during $\gamma$-ray flares favouring a smaller angle. However, we cannot rule out that PKS 0521$-$36 is the misaligned counterpart of a synchrotron-dominated blazar.
\end{abstract}


\section{Introduction}

Active galactic nuclei (AGN) are compact regions at the centre of a few percent of galaxies with a non-stellar emission overwhelming the
thermal contribution of the entire galaxy. AGN include a super-massive black hole (SMBH) as central engine, whose strong gravitational
potential pulls the surrounding materials inwards, forming a disc of hot
plasma. In addition, gas clouds move in the potential well of the SMBH,
producing optical and UV emission lines. The central region is surrounded by absorbing material in a flattened configuration, idealized as a toroidal shape, located at $\sim$1--10 pc. In radio-loud objects there is the
additional presence of relativistic jets, roughly perpendicular to the
disc. According to the Unified Model \citep{urry95}, AGN types are classified on the basis of the orientation of their jet with respect to the observer. Blazars, usually divided into flat spectrum radio quasars (FSRQ) and BL Lac objects, represent the fraction of AGN with their jet oriented at very small viewing angle, which causes relativistic aberration and emission amplification \citep{blandford78}.
Radio galaxies should be the misaligned parent population of blazars. On the basis of their radio
morphology and power, radio galaxies are classified as Fanaroff Riley type I
(FR I) or Fanaroff Riley type II (FR II). Decelerating jets and kpc scale
edge-darkened lobes are found in the weaker FR I radio
galaxies, while relativistic jets and edge-brightened radio lobes are found in the stronger
FR II radio galaxies \citep{fanaroff74}. According to the Unification scenario for radio-loud AGN, FR I radio galaxies correspond to BL Lac objects, and FR II radio galaxies are the parent population of FSRQ.

PKS 0521$-$36 was first classified as an N galaxy \citep{bolton65}, and then a
BL Lac object \citep{danziger79, burbidge87}. However, this source shows broad
and variable nuclear emission lines in optical and UV bands \citep{ulrich81,
  scarpa95} with equivalent width (EW) much larger than 5 \AA\, (rest-frame),
the threshold historically proposed to distinguish between BL Lac objects
and FSRQ \citep[e.g.,][]{stickel91,stocke91}. This suggests
that PKS 0521$-$36 is a misclassified AGN. For this source, a large-scale optical jet well aligned with the kiloparsec radio jet was observed, with a
clear correspondence between the radio and optical structures \citep[and references
therein]{scarpa99}. In radio and optical bands the jet resembles that of the nearby
radio galaxy M\,87 \citep[e.g.,][]{sparks94}. {\em Chandra} detected the X-ray counterpart of the
innermost 2-arcsec jet \citep{birkinshaw02}. 

In the $\gamma$-ray energy range PKS 0521$-$36 was tentatively detected by the Energetic Gamma Ray Experiment Telescope (EGRET) on board
the {\em Compton Gamma-Ray Observatory} in Phase 1 \citep{lin95}, but it was
not included in the Third EGRET catalogue \citep{Hartman99}. No beaming effect
is needed for the core brightness temperature estimated in radio, consistent
with the non-detection of superluminal motion
\citep{tingay02}.
By modeling the spectral energy distribution (SED) and taking into account its
radio characteristics, \citet{pian96} derived a viewing angle of 30$^{\circ}$
with a bulk Lorentz factor of 1.2. Very Long Baseline Array (VLBA) image showed
that the same Position Angle (PA) found on the parsec-scale jet is
maintained, without any significant bending, over 3 orders of magnitude of length scale
\citep{giroletti04}. This may be a further indication of a relatively large
viewing angle. For a plausible Lorentz factor of $\Gamma$ = 5,
\citet{giroletti04} derive a viewing angle $\theta$ in the range 21$^{\circ}$ $-$ 27$^{\circ}$.

On 2010 June 17, a strong $\gamma$-ray flare from PKS 0521$-$36 was detected by the
Large Area Telescope (LAT) on board the {\em Fermi Gamma-ray Space Telescope} satellite \citep{iafrate10}, triggering a {\em Swift} follow-up observation that
confirmed the high activity of the source in optical, UV, and X-rays \citep{dammando10}.
In this paper we investigate the nature of this object and its emission
mechanisms by the analysis of multifrequency data
collected from radio to $\gamma$ rays, focusing in particular on the 2010
June flaring activity. 

The paper is organized as follows: in Section 2, we report the LAT data analysis and results. In Section
3, we report the results of the new {\em Swift} and archival {\em XMM--Newton}
and {\em Chandra} data analysis. Radio data
collected by the VLBA, Very Large Array (VLA), University of Michigan Radio
Astronomy Observatory (UMRAO), and Submillimeter Array (SMA) are presented in
Section 4. In Section 5, we discuss the source properties from radio to $\gamma$ rays. In Section 6, we discuss
the modeling of the overall SED and draw our conclusions in Section 7. 

Throughout the paper the quoted uncertainties are given
at the 1$\sigma$ level, unless otherwise stated. The photon indices are parameterized as $dN/dE \propto E^{-\Gamma_{\nu}}$, where
  $\Gamma_{\nu}$ is the photon index at the different energy bands. We used a
  $\Lambda$ cold dark matter cosmology with H$_0$ = 71 km
  s$^{-1}$ Mpc$^{-1}$, $\Omega_{\Lambda}$ = 0.73, and $\Omega_{m}$ = 0.27 \citep{komatsu09}. The
corresponding luminosity distance at $z$ = 0.056 is d$_{\rm\,L}$ = 246.8 Mpc and 1
arcsec corresponds to a projected size of 1.073 kpc.

\section{{\em Fermi}-LAT Data}
\label{FermiData}

The LAT on board the {\em Fermi} satellite is a $\gamma$-ray telescope operating from $20$\,MeV to $>300$\,GeV, with a
large peak effective area ($\sim$ $8000$\,cm$^2$ for $1$\,GeV photons), an energy resolution typically $\sim$10 per cent, and a field of view of about 2.4\,sr with single-photon angular resolution (68 per cent containment radius) of 0\fdg6 at {\em E} = 1 GeV on-axis. Details about the LAT are given by \citet{atwood09}.

\begin{figure}
\centering
\includegraphics[width=7cm]{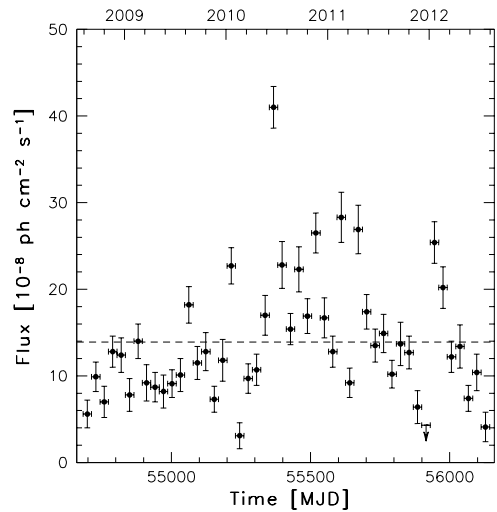}
\caption{LAT light curve of PKS 0521$-$36 for E $>$ 100 MeV using a PL model with photon index fixed to $\Gamma_{\gamma}$ = 2.45 in monthly time bins from 2008 August 4 to 2012 August 4 (MJD 54682--56143). The downward arrow represents a 2$\sigma$ upper limit. The dashed line indicates the average flux over the whole period.}
\label{Fig1}
\end{figure}

The LAT data reported in this paper were collected over the first 4 years of
{\em Fermi} operation, from 2008 August 4 (MJD 54682) to 2012 August 4 (MJD
56143) in the 0.1--100 GeV energy range. During this time, the {\em Fermi} observatory operated almost entirely
in survey mode. The analysis was performed with the \texttt{ScienceTools}
software package version v9r32p5 \footnote{http://fermi.gsfc.nasa.gov/ssc/data/analysis/}.
Only events belonging to the `Source' class were used. The time intervals when
the rocking angle of the LAT was greater than 52$^{\circ}$ were rejected. In
addition, a cut on the zenith angle ($< 100^{\circ}$) was applied to reduce
contamination from the Earth limb $\gamma$ rays, which are produced by cosmic
rays interacting with the upper atmosphere. The spectral analysis was
performed with the instrument response functions \texttt{P7REP\_SOURCE\_V15}
using an unbinned maximum-likelihood method implemented  in the tool
\texttt{gtlike}. Isotropic (`iso\_source\_v05.txt') and Galactic diffuse
emission (`gll\_iem\_v05\_rev1.fit') components were used to model the background\footnote{http://fermi.gsfc.nasa.gov/ssc/data/access/lat/Background\\Models.html}. The normalizations of both components were allowed to vary freely during the spectral fitting. 

\begin{figure}
\centering
\includegraphics[width=7cm]{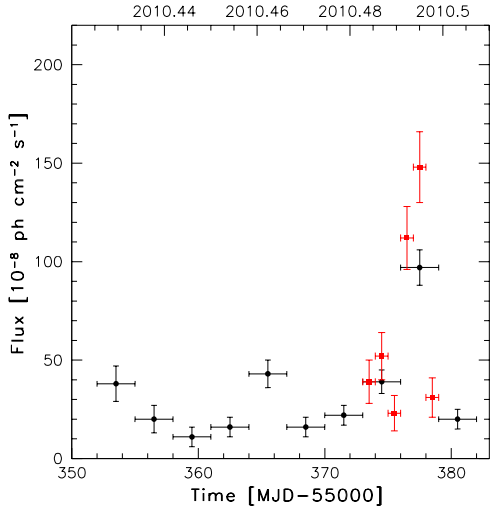}
\caption{LAT light curve of PKS 0521$-$36 for E $>$ 100 MeV using a PL model with photon index
  fixed to $\Gamma_{\gamma}$ = 2.16 in 3-day (black circles) and 1-day (red squares) time bins from 2010 June 4 to July 4 (MJD 55351--55381).}
\label{Fig2}
\end{figure}

\begin{table*}
\caption{Unbinned likelihood spectral fit results.}
\begin{tabular}{lccc|cccc}
 \hline 
&  &  \multicolumn{2}{c}{PL} & \multicolumn{3}{c}{LP} \\
Time Period (MJD)& Time Period (UT) & $\Gamma$ & TS$_{\rm PL}$ & $\alpha$
&$\beta$  & TS$_{\rm LP}$ \\
\hline
54682--56143 & 2008-08-04/2012-08-04 & 2.45$\pm$0.02 & 4945 & 2.40$\pm$0.05 & 0.07$\pm$0.02 & 4962\\       
55351--55381 & 2010-06-04/2010-07-04 & 2.16$\pm$0.05 &  901 & 2.14$\pm$0.07 & 0.02$\pm$0.02 & 902\\
\hline
  \end{tabular}
\label{LAT}
\end{table*} 

We analysed a region of interest of $10^{\circ}$ radius centred at the
location of PKS 0521$-$36. We evaluated the significance of the $\gamma$-ray signal from the source by
means of the maximum-likelihood test statistic TS = 2$\times$(log$L_1$ - log$L_0$), where
$L$ is the likelihood of the data given the model with ($L_1$) or without
($L_0$) a point source at the position of PKS 0521$-$36
\citep[e.g.,][]{mattox96}. The source model used in \texttt{gtlike} includes
all of the point sources from the third {\em Fermi}-LAT catalogue \citep[3FGL;][]{acero15} that fall within $15^{\circ}$ of
the source. The spectra of these sources were parametrized by  power-law (PL),
log-parabola (LP), or exponentially cut-off power-law model, as in the 3FGL
catalogue.
A first maximum-likelihood analysis was performed to remove from the model
faint sources with fluxes lower than 1$\times$10$^{-8}$ ph cm$^{-2}$ s$^{-1}$
for avoiding possible problems of fit convergence. 
A second maximum-likelihood analysis
was performed on the updated source model. In the fitting procedure, the
normalization factors and the spectral parameters of
the sources lying within 10$^{\circ}$ of PKS 0521$-$36 were left as free
parameters. For the sources located between 10$^{\circ}$ and 15$^{\circ}$ from
our target, we kept the normalization and the spectral parameters fixed to the values
from the 3FGL catalogue.

\begin{figure}
\centering
\includegraphics[width=7cm]{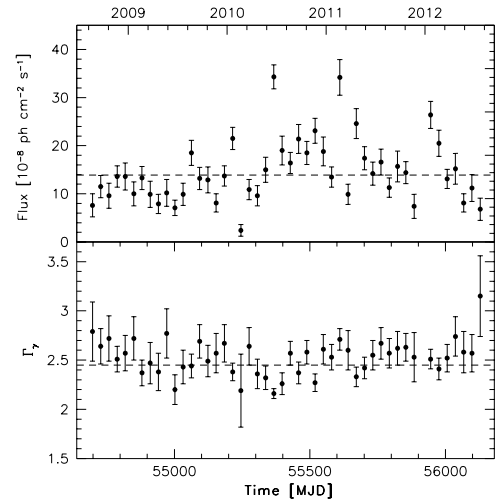}
\caption{The 0.1--100 GeV flux in units of 10$^{-8}$ ph cm$^{-2}$ s$^{-1}$ (top
  panel) and the photon index from a PL model (bottom panel) for PKS 0521$-$36 using 1-month time-bins. The dashed line in both panels represents the mean value.}
\label{Fig3}
\end{figure}

\begin{figure}
\centering
\includegraphics[width=7cm]{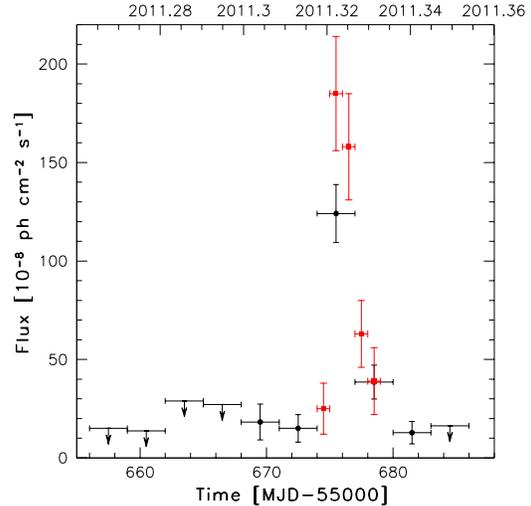}
\caption{LAT light curve of PKS 0521$-$36 for E $>$ 100 MeV using a PL model with photon index
  fixed to $\Gamma_{\gamma}$ = 2.33 in 3-day (black circles) and 1-day (red squares)
  time bins from 2011 April 6 to May 5 (MJD 55657--55686). The downward arrow represents a 2$\sigma$ upper limit.}
\label{Fig4}
\end{figure}

The fit with a PL model, $dN/dE \propto$ $(E/E_{0})^{-\Gamma_{\gamma}}$, to the data integrated over 48
months of {\em Fermi} operation (2008 August--2012 August) in the 0.1--100 GeV energy range results in TS = 4945, with an
integrated average flux of (13.9 $\pm$ 0.4)$\times$10$^{-8}$ ph cm$^{-2}$ s$^{-1}$, and a photon index $\Gamma_{\gamma}$ = 2.45 $\pm$ 0.02. Figure~\ref{Fig1} shows the $\gamma$-ray light curve of the first 4 years of
{\em Fermi} observations of PKS 0521$-$36 built using 1-month time bins. For each time bin, the photon index was frozen to the value
resulting from the likelihood analysis over the whole period. The systematic
uncertainty in the flux is dominated by the systematic uncertainty in the
effective area \citep{ackermann12b}, which amounts to 10 per cent below 100
MeV, decreasing linearly with the logarithm of energy to 5 per cent between 316 MeV and 10 GeV, and increasing linearly with the logarithm
of energy up to 15 per cent at 1 TeV \footnote{http://fermi.gsfc.nasa.gov/ssc/data/analysis/LAT\_caveats.html}. 

PKS 0521$-$36 has been quite active since 2009 September. An increase of the
$\gamma$-ray flux was observed in 2010 January, and subsequently a significant
flaring activity occurred in 2010 June. A second light curve focusing on the
period of the highest activity (2010 June 4--July 4; MJD 55351--55381) was
built with 3-day and 1-day
time bins (Fig.~\ref{Fig2}). Leaving the photon index free to vary during 2010
June 4--July 4, the global fit results in TS = 901 and a photon index $\Gamma_{\gamma}$ = 2.16 $\pm$ 0.05, indicating a
harder-when-brighter behaviour, similar to what was observed in FSRQs and
low-synchrotron-peaked BL Lacs \citep{abdo10b}. The peak of the emission was
observed on 2010 June 30 (MJD 55377), with a flux of (148 $\pm$ 18)$\times$10$^{-8}$ ph cm$^{-2}$ s$^{-1}$ in the 0.1--100 GeV energy range,
a factor of $\sim$11 higher than the average $\gamma$-ray flux during 2008 August--2012 August. The corresponding apparent isotropic
$\gamma$-ray peak luminosity is 8.5$\times$10$^{45}$ erg
s$^{-1}$. During the period 2010 June 4--July 4,  the
fit with an LP model, $dN/dE \propto$ $(E/E_{0})^{-\alpha-\beta \, \log(E/E_0)}$, in the 0.1--100 GeV energy range results in TS
= 902, with a spectral slope $\alpha$ = 2.14 $\pm$ 0.07 at the decorrelation energy $E_0$ = 384 MeV, and a curvature parameter $\beta$ = 0.02
$\pm$ 0.02 (Table~\ref{LAT}). We used a likelihood ratio test (LRT) to check a
PL model (null hypothesis) against an LP model (alternative hypothesis). Following \citet{nolan12}, these values may be compared by defining the curvature test
statistic TS$_{\rm curve}$=(TS$_{\rm LP}$ - TS$_{\rm PL}$). The LRT results in TS$_{\rm curve}$ = 1, corresponding to a $\sim$1$\sigma$ difference,
indicating no significant curvature in the $\gamma$-ray spectrum during the
flaring period. Considering the entire period 2008 August 4 -- 2012 August 4
the fit with an LP results in TS = 4962, with a spectral slope $\alpha$ = 2.40
$\pm$ 0.05 at the decorrelation energy $E_0$ = 384 MeV, a curvature parameter $\beta$ = 0.07
$\pm$ 0.02, and an average flux of (13.2 $\pm$ 0.8)$\times$10$^{-8}$ ph
cm$^{-2}$ s$^{-1}$ (Table~\ref{LAT}), consistent with the results reported
in the 3FGL \citep{acero15}. The LRT results in TS$_{\rm curve}$ = 17, corresponding to a
$\sim$4$\sigma$ difference (for 1 degree of freedom). This indicates a significant curvature in the
$\gamma$-ray spectrum over the entire period, contrary to what is observed in shorter periods.

\noindent In addition to the 3-day and 1-day light curves shown in
Fig.~\ref{Fig2}, we also computed a light curve in 12-hour bins during the
period of brightest flux (2010 June 26--July 1). In this light curve (not shown) we note a significant increase from (49 $\pm$ 15)$\times$10$^{-8}$ ph cm$^{-2}$ s$^{-1}$ to (193 $\pm$ 29)$\times$10$^{-8}$ ph cm$^{-2}$ s$^{-1}$ between the first
and second 12-hr bin on 2010 June 29. The peak flux at 12-hr time-scale
corresponds to an apparent isotropic $\gamma$-ray luminosity of 1.1$\times$10$^{46}$ erg s$^{-1}$.

We investigated whether spectral changes are present during the period
2008 August--2012 August, using a PL model. We plot the photon index against the $\gamma$-ray flux above 100
MeV estimated on a monthly time-scale. Unlikely as in Fig.~\ref{Fig1}, in which the photon indices are fixed, for each time bin the spectral parameters for PKS 0521$-$36 and for all the sources within 10$^{\circ}$ from it were left
free to vary. No obvious relation between flux and photon index was observed (Fig.~\ref{Fig3}).

Other high-activity periods of PKS 0521$-$36 have been observed in 2011 February and April, and 2012 January. In particular, in 2011
April the source was not detected for the first part of the month, and then a
rapid increase of flux from (25 $\pm$ 13)$\times$10$^{-8}$ ph cm$^{-2}$
s$^{-1}$ to (185 $\pm$ 29)$\times$10$^{-8}$ ph cm$^{-2}$ s$^{-1}$ was observed between
2011 April 23 and 24 (Fig.~\ref{Fig4}) with $\Gamma_{\gamma}$ = 2.33, the value obtained leaving the photon index free to
vary on monthly time-scales (see Fig.~\ref{Fig3}, bottom panel). As
for Fig.~\ref{Fig2}, we investigated the 12-hr light curve during the period
of brightest flux (2011 April 22--26). We note that an increase from (129 $\pm$ 32)$\times$10$^{-8}$ ph cm$^{-2}$
s$^{-1}$ to (275 $\pm$ 56)$\times$10$^{-8}$ ph cm$^{-2}$ s$^{-1}$ was observed
between the first and second 12-hr bin on 2011 April 24. The peak flux at 12-hr time-scale
corresponds to an apparent isotropic $\gamma$-ray luminosity of
1.2$\times$10$^{46}$ erg s$^{-1}$. It is interesting to note that during the 2010
June and 2011 April flares the peaks show a moderately asymmetric profile (i.e. different
rising and decaying times) on a daily time scale. In particular, the rapid
increase observed in 2011 April suggests a cooling time longer than the light
crossing time R/$c$ related to fast injection of accelerated particles and a slower radiative cooling.

Analyzing the LAT events with $E > 10$ GeV collected over 2008 August--2012
August the fit results in TS = 89, a photon index $\Gamma_{\gamma}$ = 2.71
$\pm$ 0.27, and a flux of (1.43 $\pm$ 0.36)$\times$10$^{-10}$ ph cm$^{-2}$
s$^{-1}$. By means of the \texttt{gtsrcprob} tool, we estimated that the
highest energy photon emitted from PKS 0521$-$36 (with probability $>$ 90 per cent of being
associated with the source) was observed on 2010 August 17 (MJD 55425), at a
distance of 0\fdg03 from the source and with an energy of 73 GeV.

\section{X-ray, UV, and optical data} 

\subsection{{\em Swift} observations}
\label{SwiftData}

The {\em Swift} satellite \citep{gehrels04} performed nineteen observations of
PKS 0521$-$36 between 2005 May 26 and 2011 September 5. In particular five
observations were triggered by the $\gamma$-ray activity observed in 2010
June. The observations were performed with all three on-board instruments: the X-ray Telescope \citep[XRT;][0.2--10.0 keV]{burrows05}, the UV/Optical Telescope \citep[UVOT;][170--600 nm]{roming05}, and the Burst Alert Telescope \citep[BAT;][15--150 keV]{barthelmy05}. 

\subsubsection{Swift-BAT}

The hard X-ray flux of this source is below the sensitivity  of the BAT
instrument for the short exposure time of the individual {\em Swift}
observations. However, PKS 0521$-$36 is detected in the BAT 70-month catalogue\footnote{http://heasarc.gsfc.nasa.gov/results/bs70mon}, generated from the all-sky survey in the time period 2004 November--2010 August. The data reduction and extraction procedure of the 8-channel spectrum is described in \citet{baumgartner13}. The 14--195 keV spectrum is described by a power law
with photon index of 1.83 $\pm$ 0.15 ($\chi$$^2_{\rm red}$ = 2.87, 5 d.o.f). The resulting 14--195 keV flux is (3.5 $\pm$ 0.3)$\times$10$^{-11}$ erg cm$^{-2}$ s$^{-1}$.

\begin{table*}
\caption{Log and fitting results of {\em Swift}-XRT observations of PKS
  0521$-$36. A power-law model with $N_{\rm H}$ fixed to the Galactic column density was used.}
\label{XRT}           
\centering                          
\begin{tabular}{c c c c c c}       
\hline
Date &  Date & Net Exposure Time & $\Gamma_{\rm X}$ &  Flux (2--10 keV) &
$\chi^2_{\rm red}$ (d.o.f.)\\             
(MJD) & (UT) & (s)   &  & (10$^{-11}$ erg cm$^{-2}$ s$^{-1}$) & \\    
\hline                        
53516 & 2005-26-05 & 900  & 1.72$\pm$0.14 & 1.34$\pm$0.12 & 1.034 (17)\\
54498 & 2008-02-02 & 4832 & 1.61$\pm$0.06 & 1.32$\pm$0.06 & 1.062 (88)\\
54503/04 & 2008-02-07/08 & 4947 & 1.59$\pm$0.06 & 1.18$\pm$0.06 & 0.918 (64)\\
54509 & 2008-02-13 & 2901 & 1.55$\pm$0.09 & 1.27$\pm$0.08 & 0.869 (41)\\
55260 & 2010-03-05 & 1681 & 1.67$\pm$0.12 & 0.88$\pm$0.11 & 1.082 (20)\\
55263 & 2010-03-08 & 2405 & 1.62$\pm$0.10 & 1.20$\pm$0.10 & 0.887 (35)\\
55366 & 2010-06-19 & 2759 & 1.62$\pm$0.08 & 1.86$\pm$0.10 & 1.018 (42)\\
55370 & 2010-06-23 & 3951 & 1.57$\pm$0.07 & 2.38$\pm$0.14 & 1.020 (57)\\
55382 & 2010-07-05 & 2774 & 1.64$\pm$0.09 & 1.97$\pm$0.13 & 1.089 (42)\\
55386 & 2010-07-09 & 2904 & 1.66$\pm$0.09 & 2.06$\pm$0.15 & 1.073 (37)\\
55390 & 2010-07-13 & 2822 & 1.60$\pm$0.08 & 1.90$\pm$0.11 & 0.992 (40)\\
55626 & 2011-03-06 & 1126 & 1.78$\pm$0.16 & 1.30$\pm$0.18 & 0.776 (13)\\
55627 & 2011-03-07 & 2015 & 1.75$\pm$0.10 & 1.49$\pm$0.09 & 0.899 (32)\\
55661/02 & 2011-04-09/10 & 1868 & 1.62$\pm$0.08 & 1.34$\pm$0.09 & 0.968 (34)\\
55804/05 & 2011-08-31/09-01 & 1628 & 1.79$\pm$0.12 & 0.80$\pm$0.08 & 1.117 (17) \\
55809 & 2011-09-05 & 1393 & 1.67$\pm$0.12 & 1.77$\pm$0.15 & 0.928 (21) \\
\hline                               
\end{tabular}
\end{table*}

\subsubsection{Swift-XRT}

The XRT data were processed with standard procedures (\texttt{xrtpipeline v0.12.6}), filtering, and screening criteria by using the \texttt{HEAsoft} package (v6.13). The data were collected in photon counting mode in all the observations, and only XRT event grades 0--12 were selected.
Source events were extracted from a circular region with a radius of 20 pixels
(1 pixel = 2.36 arcsec), while background events were extracted from a
circular region with radius of 50 pixels away from the source region. Some
observations showed a source count rate $>$ 0.5 counts s$^{-1}$; thus pile-up
correction was required. For those observations we extracted the source events
from an annular region with an inner radius of 3 pixels (estimated by means of
the PSF fitting technique) and an outer radius of 30 pixels, while background
events were extracted from an annular region centered on the source with radii
of 70 and 120 pixels. Ancillary response files were generated with
\texttt{xrtmkarf}, and account for different extraction regions, vignetting
and PSF corrections. We used the spectral redistribution matrices in the
calibration database maintained by HEASARC\footnote{http://heasarc.gsfc.nasa.gov/}. The data collected during 2008 February 7 and 8, 2011 April 9 and 10, and 2011 August 31 and September 1 were summed in order to have enough statistics to obtain a good spectral fit. 

\begin{figure}
\includegraphics[width=7.5cm]{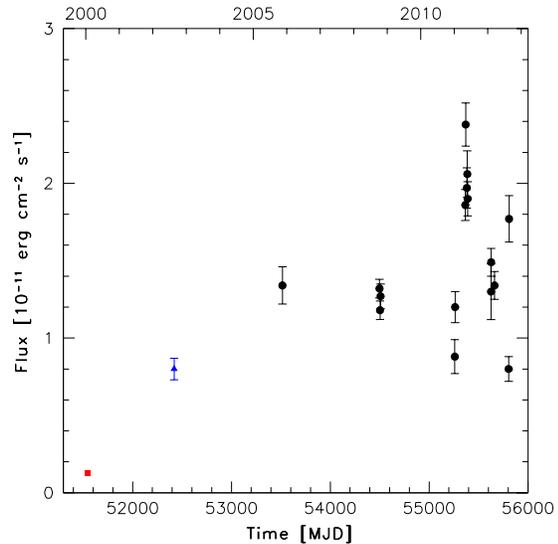}
\caption{X-ray light curve in the 2--10 keV energy range during the period 1999--2012. Square: {\em Chandra}; triangle: {\em XMM--Newton}; circles: {\em Swift}-XRT.}
\label{X-ray}
\end{figure}

We fit the spectra in the 0.3--10 keV energy range with an absorbed power law using
the photoelectric absorption model \texttt{tbabs} \citep{wilms00}, with a neutral
hydrogen column density fixed to its Galactic value \citep[3.58$\times$10$^{20}$
cm$^{-2}$;][]{kalberla05}. Considering the lower statistics with respect to
the {\em XMM--Newton} and {\em Chandra} observations, more detailed spectral modelling was not performed with the XRT observations.
The fit results are reported in
Table~\ref{XRT}. {\em Swift}-XRT observed the source during 2005--2011 with a 2--10 keV flux in the range
(0.8--2.4)$\times$10$^{-11}$ erg cm$^{-2}$ s$^{-1}$. No significant change of
the photon index was observed. During 2010 mid-June, in the period of highest $\gamma$-ray
activity, an increase of the X-ray flux was observed, but not accompanied by
significant hardening of the spectrum. Furthermore, by comparing the 2--10
keV flux observed by {\em Swift}-XRT with those observed by {\em Chandra} (see Sect.~\ref{chandraxmm} for details) we noted a significant increase
of the flux (a factor of $\sim$10) between 1999 and 2011 (Fig.~\ref{X-ray}).

\subsubsection{Swift-UVOT}

UVOT observed PKS 0521$-$36 in all its optical ($v$, $b$, and $u$) and UV
($uvw1$, $uvm2$, and $uvw2$) photometric filters. Data were reduced with the \texttt{HEAsoft} package v6.14 and the 20130118 CALDB-UVOTA
release. We extracted the source counts from a circle with 5 arcsec radius
centred on the source. Background counts were derived from an annular region centred on the source with 15 and 20 arcsec radii. The observed magnitudes are reported in
Table~\ref{UVOT}. During 2008--2011 the difference between the maximum and minimum
magnitude is 0.7, 0.9, 0.7, 1.3, 1.3 and 1.4 mag from the $v$ to the $uvw2$
band, with the peak of the activity observed on 2010 June 19.

As in e.g. \citet{raiteri11}, we calculated the effective
wavelengths, count-to-flux conversion factors ($\rm CF_\Lambda$), and amount
of Galactic extinction in the UVOT bands ($\rm A_\Lambda$) by convolving the physical quantities
with a power-law fit to the source flux and with the UVOT effective areas. In
particular, to derive the $\rm A_\Lambda$ we adopted the mean extinction law
by \citet{cardelli89} with $R_{V}$ = $A_{V}$/$E(B-V)$ = 3.1 (the
standard value for the diffuse interstellar medium), as well as $A_{V}$ =
0.112 from \citet{schlafly11}. To obtain the UVOT de-reddened fluxes we
multiplied the count rates by the $\rm CF_\Lambda$ and corrected for the
corresponding Galactic extinction values $\rm A_\Lambda$.

\begin{table*}
\caption{Results of {\em Swift}-UVOT observations of PKS 0521$-$36 in observed magnitude.}
\label{UVOT}           
\centering                          
\begin{tabular}{c c c c c c c c}       
\hline
Date &  Date & $v$ & $b$ & $u$ & $w1$ & $m2$ & $w2$ \\             
(MJD) & (UT) & (mag)  & (mag) & (mag) & (mag) & (mag) & (mag) \\    
\hline                        
53516 & 2005-26-05 & --  & -- & -- & 15.94$\pm$0.02 & -- & -- \\
54498 & 2008-02-02 & --  & -- & -- & 15.11$\pm$0.02 & -- & -- \\
54503 & 2008-02-07 & --  & -- & -- & 15.67$\pm$0.04 & -- & -- \\
54504 & 2008-02-08 & 15.16$\pm$0.03 & 15.89$\pm$0.03 & 15.41$\pm$0.03 & 15.58$\pm$0.03 & 15.71$\pm$0.03 & 15.82$\pm$0.03 \\
54509 & 2008-02-13 & 15.15$\pm$0.03 & 15.93$\pm$0.03 & 15.42$\pm$0.03 & 15.59$\pm$0.03 & 15.69$\pm$0.04 & 15.83$\pm$0.03 \\
55260 & 2010-03-05 & 14.94$\pm$0.04 & 15.68$\pm$0.03 & 15.13$\pm$0.03 & 15.34$\pm$0.03 & 15.44$\pm$0.04 & 15.59$\pm$0.03 \\
55263 & 2010-03-08 & -- &  -- & --  & 15.18$\pm$0.02 & -- & -- \\
55366 & 2010-06-19 & 14.58$\pm$0.03 & 15.12$\pm$0.02 & 14.49$\pm$0.03 & 14.54$\pm$0.02 & 14.63$\pm$0.03 & 14.54$\pm$0.03 \\
55370 & 2010-06-23 & 14.98$\pm$0.03 & 15.67$\pm$0.03 & 15.18$\pm$0.03 & 15.36$\pm$0.03 & 15.51$\pm$0.03 & 15.63$\pm$0.03 \\
55382 & 2010-07-05 & 14.83$\pm$0.03 & 15.48$\pm$0.03 & 14.93$\pm$0.03 & 15.11$\pm$0.03 & 15.17$\pm$0.03 & 15.27$\pm$0.03 \\
55386 & 2010-07-09 & 14.93$\pm$0.03 & 15.61$\pm$0.03 & 15.08$\pm$0.03 & 15.28$\pm$0.03 & 15.33$\pm$0.03 & 15.46$\pm$0.03 \\
55390 & 2010-07-13 & 15.09$\pm$0.03 & 15.87$\pm$0.03 & 15.33$\pm$0.03 & 15.44$\pm$0.03 & 15.67$\pm$0.04 & 15.74$\pm$0.03 \\
55626 & 2011-03-06 & 15.27$\pm$0.05 & 16.05$\pm$0.04 & 15.58$\pm$0.04 & 15.71$\pm$0.04 & 15.90$\pm$0.05 & 15.97$\pm$0.04 \\
55627 & 2011-03-07 & -- & -- & -- & 15.74$\pm$0.02 & -- & -- \\
55661 & 2011-04-09 & 15.14$\pm$0.05 & 15.95$\pm$0.04 & 15.50$\pm$0.05 & 15.59$\pm$0.05 & 15.76$\pm$0.06 & 15.83$\pm$0.04 \\
55662 & 2011-04-10 & 15.10$\pm$0.05 & 15.90$\pm$0.04 & 15.39$\pm$0.04 & 15.56$\pm$0.05 & 15.69$\pm$0.05 & 15.83$\pm$0.04 \\
55804 & 2011-08-31 & -- & -- & 15.63$\pm$0.03 & -- & -- & --\\
55805 & 2011-09-01 & 15.23$\pm$0.05 & 15.99$\pm$0.04 & 15.53$\pm$0.04 & 15.81$\pm$0.05 & 15.70$\pm$0.05 & 15.85$\pm$0.03 \\
55809 & 2011-09-05 & 14.94$\pm$0.04 & 15.57$\pm$0.03 & 15.13$\pm$0.04 & 15.39$\pm$0.04 & 15.58$\pm$0.04 & 15.78$\pm$0.03 \\
\hline                               
\end{tabular}
\end{table*}

\subsection{{\em XMM--Newton} and {\em Chandra} observations}
\label{chandraxmm}

\begin{table*}
\begin{center}
\caption{Spectral results of the nucleus and the jet of PKS 0521$-$36 for the {\em Chandra} observation performed on 1999 December 31. A power-law model was used.}
\begin{tabular}{l l c}
\hline Component & Parameter & Value \\ 
\hline Nucleus & N$_{\rm H}$ {\small (cm$^{-2}$)} & (4.82 $\pm$ 0.03)$\times$10$^{20}$ \\
               & $\Gamma$ & 1.7 $\pm$ 0.1 \\
               & norm {\small (ph~keV$^{-1}$~cm$^{-2}$~s$^{-1}$)} &(3.1 $\pm$ 0.4)$\times$10$^{-4}$   \\
               & F$_{\rm 2-10\,keV}$ {\small (erg~cm$^{-2}$~s$^{-1}$)}  & (1.3 $\pm$ 0.2)$\times$10$^{-12}$ \\
\hline Jet     & N$_{\rm H}$ {\small (cm$^{-2}$)} & 3.58$\times$10$^{20}$ (fixed) \\
               & $\Gamma$ & 2.2 $\pm$ 0.3  \\
               & norm {\small (ph~keV$^{-1}$~cm$^{-2}$~s$^{-1}$)} &(2.4$^{+1.2}_{-0.3}$)$\times$10$^{-5}$   \\
               & F$_{\rm 2-10\,keV}$ {\small (erg~cm$^{-2}$~s$^{-1}$)}  &  (6.0$^{+3.0}_{-1.0}$)$\times$10$^{-14}$ \\
\hline
\end{tabular}
\label{tab_chandra}
\end{center}
\end{table*}

PKS~0521--36 was observed by the {\em Chandra} and {\em XMM--Newton} X--ray
satellites on 1999 December 31 (MJD 51543) and 2002 May 26 (MJD 52420), respectively. 
The spectral analysis was performed using the \texttt{XSPEC} v12.7 package.\\

{\em Chandra} ACIS--S spectra and instrument responses were generated using
the Chandra Interactive Analysis of Observations (CIAO) v4.3 and the related
{\em Chandra} calibration database.  The 0.3--7~keV image shows the presence of an unresolved core, a diffuse halo
and a jet feature coincident with the inner radio/optical jet \citep[see][]{birkinshaw02}.

The nuclear spectrum was extracted from an annular region (r$_{\rm in}$ = 0.3
arcsec, r$_{\rm out}$ = 1.2 arcsec) in order to minimize pile up
effects\footnote{Using the {\small PIMMS} software (http://cxc.harvard.edu/toolkit/\\pimms.jsp) the pile up for this source was
  estimated to be around 4 per cent}. The background counts were extracted
from three source--free circular regions of r = 1.2 arcsec each. 
The 0.5--7~keV spectrum is well fitted ($\chi^{2}$/dof=187/199) by a power law absorbed by a column
density slightly in excess of the Galactic value, N$_{\rm H}$ =
(4.82 $\pm$ 0.03)$\times$10$^{20}$~cm$^{-2}$.

The spectrum of the jet was extracted from a circle of radius 1 arcsec at a
distance of 1.2 arcsec from the core. The background was chosen from three
source--free circular regions. The best fit model is a power law with photon
index $\Gamma_{\rm X}$ = 2.2 $\pm$ 0.3. The 2--10~keV flux of the
jet is roughly twenty times lower than the core flux. {\em Chandra} spectral results for the nucleus and the jet are listed in Table~\ref{tab_chandra}.

The {\em XMM--Newton} EPIC observation of PKS~0521$-$36 was analyzed using the Science Analysis System (SAS) v11.0 software and
available calibration files. Time intervals affected by high background were excluded. After this data cleaning, we obtained a net
exposure of 26.6~ks and a count rate of 3.50 $\pm$ 0.01 count~s$^{-1}$ for the
pn. The source and background spectra were extracted from circular regions of
32 arcsec and 44 arcsec radius, respectively. The response matrices were
created using the {\small SAS} commands {\small RMFGEN} and {\small ARFGEN}. The nuclear data are not piled up. Data were grouped into 25 counts per bin in order to apply the $\chi^{2}$ statistic. A power law, absorbed by Galactic column density, plus a thermal ({\small
  APEC}) component is a good parametrization of the data
($\chi^{2}$/dof=972/925). In \citet{foschini06}, a broken power law model
absorbed by the Galactic column was reported as the best fit to these {\em XMM--Newton} data.
However the 0.5--10~keV best fit model turns out to
be a combination of a broken power-law, describing the nuclear radiation, plus a
Raymond--Smith component ($\chi^{2}$/dof=947/923), which accounts for the
thermal emission of the diffuse halo around the source, necessarily inside the
32 arcsec extraction radius. It is interesting to note
  that the two spectral indices of the broken power-law are similar to those
  observed for the nucleus and the jet by {\em Chandra}. This may indicate that we are
  observing both these components in the {\em XMM--Newton} spectrum. An alternative possibility is that in the {\em XMM--Newton} spectrum we are observing the tail of the synchrotron emission and the rise of the IC component from the jet. There is no significant evidence for a Fe K$\alpha$ emission line. {\em XMM--Newton} spectral results are listed in Table~\ref{tab_xmm}.\\

\begin{table}
\begin{center}
\caption{Spectral results for the {\em XMM--Newton} observation
  of PKS 0521$-$36 performed on 2002 May 26. A broken
  power-law model with $N_{\rm H}$ fixed to the Galactic
absorption plus a Raymond-Smith component was used.} 

\begin{tabular}{l c}
\hline Parameter & Value \\ 
\hline $\Gamma_{1}$ & 2.4 $\pm$ 0.2  \\
 E$_{\rm b}$ {\small (keV)} & 0.76$^{+0.07}_{-0.04}$\\
 $\Gamma_{2}$ & 1.7 $\pm$ 0.2  \\
 norm {\small (ph~keV$^{-1}$~cm$^{-2}$~s$^{-1}$)} & (1.7 $\pm$ 0.1)$\times$10$^{-3}$   \\
 kT {\small (eV)}     & 0.85$^{+0.09}_{-0.07}$\\
 norm  {\small (ph~keV$^{-1}$~cm$^{-2}$~s$^{-1}$)} &(8.2 $\pm$ 2.7)$\times$10$^{-5}$\\
 F$_{\rm 2-10\,keV}$ {\small (erg~cm$^{-2}$~s$^{-1}$)}  & (8.0 $\pm$ 0.7)$\times$10$^{-12}$ \\
\hline
\end{tabular}
\label{tab_xmm}
\end{center}
\end{table}

\section{Radio data}

\subsection{VLBA observations}\label{vlbi}

\begin{table*}
\caption{Model-fit parameters. Columns 1 and 2: flux density $S$ of the components; columns 3 and 4: the radial distance $d$ from the core at 8 and
  15 GHz, respectively; columns 5 and 6: the polar angle $\theta$, measured
  north to east at 8 GHz and 15 GHz, respectively; columns 7 and 8: the major axis $a$ of the component ($0$ indicates a delta component); column 9: the ratio $b/a$ between the minor and major axis of the component (1 indicates a circular Gaussian); column
  10: the orientation $\phi$ of the major axis on the plane of the sky; column
  11: the spectral index $\alpha_{\rm r}$. Numbers in italics refer to
  parameters held fixed in the model fit. $^{a}$These values are left free to
vary in the model fit at 8 GHz and fixed at 15 GHz.}
\begin{center}
\begin{tabular}{ccccccccccc}
\hline
$S_{8}$ & $S_{15}$ & $d_{8}$ & $d_{15}$ & $\theta_{8}$ & $\theta_{15}$ &
$a_{8}$ & $a_{15}$ & $b/a$ & $\phi$ & $\alpha_{\rm r}$ \\ 
(mJy) & (mJy) & (mas) & (mas) & ($^{\circ}$) & ($^\circ$) & (mas) & (mas) &  & ($^{\circ}$) & \\
\hline
545 & 815 & 0.00 & 0.00 & 0.0     & 0.0     & 0.00 & 0.00 & {\it 0} & 0 & $-$0.7 \\
414 & 584 & 0.28 & 0.27 & $-$54.9 & $-$33.6 & 0.37 & 0.27 & {\it 1} & 0 & $-$0.6 \\
233 & 195 & 0.96 & 0.96 & $-$42.9 & $-$43.0 & 0.44 & 0.25 & {\it 1} & 0 & 0.3 \\
100 &  72 & 2.06 & 2.28 & $-$34.1 & $-$42.8 & 0.87 & 1.11 & {\it 1} & 0 & 0.5 \\
 33 &  12 & 4.30 & 4.35 & $-$40.8 & $-$39.9 & 1.18 & {\it 1.18} & {\it 1} & 0 & 1.7 \\
 38 &  25 & 10.8 & {\it 10.8} & $-$55.5 & {\it $-$55.5} & 4.58 & {\it 4.58} & {\it 1}   & 0 & 0.7 \\
298 & 220 & 27.9 & {\it 27.9} & $-$45.9 & {\it $-$45.9} & 7.95 & {\it 7.95} & 0.58$^{a}$ & $-$19.7$^{a}$ & 0.5 \\
 94 &  59 & 35.3 & {\it 35.3} & $-$47.3 & {\it $-$47.3} & 8.53 & {\it 8.53} & 0.47$^{a}$ & $-$6.9$^{a}$ & 0.8 \\
\hline
\end{tabular}
\label{t.modelfit}
\end{center}
\end{table*}
 
PKS~0521$-$36 was observed with the VLBA at 8 and 15 GHz on 2003 January
7. The central frequencies of the two bands were 8.421 GHz and 15.356 GHz and the
total bandwidth was 32 MHz for each band. The on-source time was about 30 and
50 min at 8 and 15 GHz, respectively, divided into 10 scans distributed
over a  wide range of hour angle, to ensure the best sampling of the
$(u,v)$-plane. One station (North Liberty) could not provide data, while two performed poorly because of snow (Hancock) and hardware problems (Brewster).

We downloaded the data from the National Radio Astronomy Observatory (NRAO)
archive and performed the typical steps of calibration in the astronomical image
processing system ({\small AIPS}). We
used system temperature and gain curve tables for the visibility amplitude
calibration and \texttt{FRING} for the delay, rate, and visibility
phase. Uncertainties on the amplitude calibration are about 10\%. Several
cycles of self-calibration were then performed in {\small DIFMAP} to improve the initial calibration, both in phase and amplitude, and produce final images.

\begin{figure}
\centering
\includegraphics[width=6.5cm]{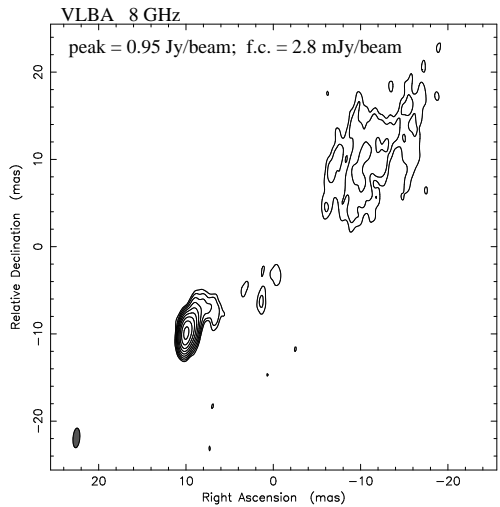}
\vspace{1cm}
\caption{VLBA image of PKS~0521$-$36 at 8 GHz. On the image we provide the
  restoring beam, plotted in the bottom-left corner, the peak flux density in
  Jy/beam, and the first contour (f.c.) intensity in mJy/beam, which is three
  times the off-source noise level. Contour levels increase by a factor of 2.\label{vlba8}}
\end{figure}

\begin{figure}
\centering
\includegraphics[width=6.5cm]{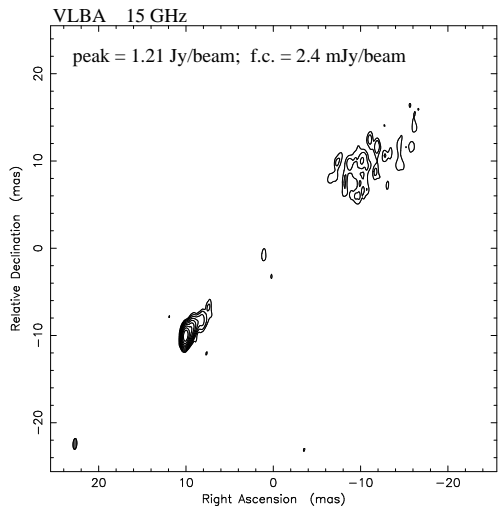}
\vspace{1cm}
\caption{VLBA image of PKS~0521$-$36 at 15 GHz. On the image we provide the restoring beam, plotted in the bottom-left corner, the peak flux density in Jy/beam, and the first contour (f.c.) intensity in mJy/beam, which is three times the off-source noise level. Contour levels increase by a factor of 2.\label{vlba15}}
\end{figure}

In Figs.~\ref{vlba8} and \ref{vlba15}, we show total intensity images of
PKS~0521$-$36 at 8 and 15 GHz, respectively. The source is well detected at
both frequencies, with a bright compact core and a one-sided jet oriented in
PA $\sim -45^{\circ}$ on the plane of the sky. The brightness profile along
the jet axis decreases rapidly with increasing distance from the core, but it
then suddenly rises again at $\sim30$ mas. The total flux density is 1.78 and
1.93~Jy at 8 and 15 GHz, respectively, with a peak brightness of 0.94 and 1.21
Jy~beam$^{-1}$. The core has an inverted spectral index of $\alpha_{\rm r}$ = --0.4$\pm0.2$ ($S(\nu)\propto\nu^{-\alpha_{\rm r}}$). The bright jet emission region at $\sim30$~mas has a flux density $S_8\sim 405$~mJy and $S_{15}\sim265$~mJy, corresponding to a spectral index of $\alpha=0.7\pm0.2$.

In Table~\ref{t.modelfit}, we report the parameters of a model fit with Gaussian components to the visibility plane. We have first model-fitted the 8 GHz visibilities with delta (for the core), circular (for the inner jet), and elliptical (for the outer blob) Gaussian components, until we reached convergence. Then, we used this model as a starting condition for the model fit to the 15 GHz visibilities. In this latter fit, we kept the position and size of the outer three components fixed, while all the other parameters (including the flux density of every component) were left free to vary. 

In order to maximize sensitivity to the diffuse jet emission, we also produced
an image applying a Gaussian taper of 0.3 at a radius of 30 M$\lambda$ to the
8 GHz dataset. This yields a resolution of $4.9\ \rm{mas} \times 14.2\ \rm{mas}$ in PA 2\fdg3, a peak brightness in the jet region of about
185~mJy~beam$^{-1}$, and an off source peak of 1.4~mJy~beam$^{-1}$ on the counter-jet side. 

\subsection{VLA data}

To obtain information on the radio flux density on kiloparsec-scale,
we retrieved archival NRAO VLA data at 4.9 and 8.4 GHz. Observations of
PKS\,0521$-$36 were performed at 4.9 GHz on 2009 August 6, when the
array was in C configuration, while 8.4-GHz observations were carried
out on 2009 October 3 when the array was in the hybrid C and D
configuration. During both runs the target was observed for about 40 s. A few
antennas did not participate in the observations due to a system upgrade. The primary flux
density calibrator was 3C\,48. Uncertainties on the amplitude calibration (1$\sigma$) are about 5 per cent at
both frequencies.
The data reduction was carried out following the standard procedures
for the VLA implemented in the NRAO {\small AIPS} package. 
The flux density at each frequency was measured on the final image
produced after a few phase-only self-calibration iterations using TVSTAT,
which performs an aperture integration over a selected region on the image
plane. Flux density of the compact components was measured 
by the {\small AIPS} task JMFIT, which performs a Gaussian fit of the source
component on the image
plane. Flux densities are reported in Table~\ref{vla_table}.
The resolution achieved is about 17 arcsec $\times$ 4 arcsec and 11 arcsec
$\times$6 arcsec at 4.9 and 8.4 GHz, respectively.   

\begin{figure}
\includegraphics[width=7.5cm]{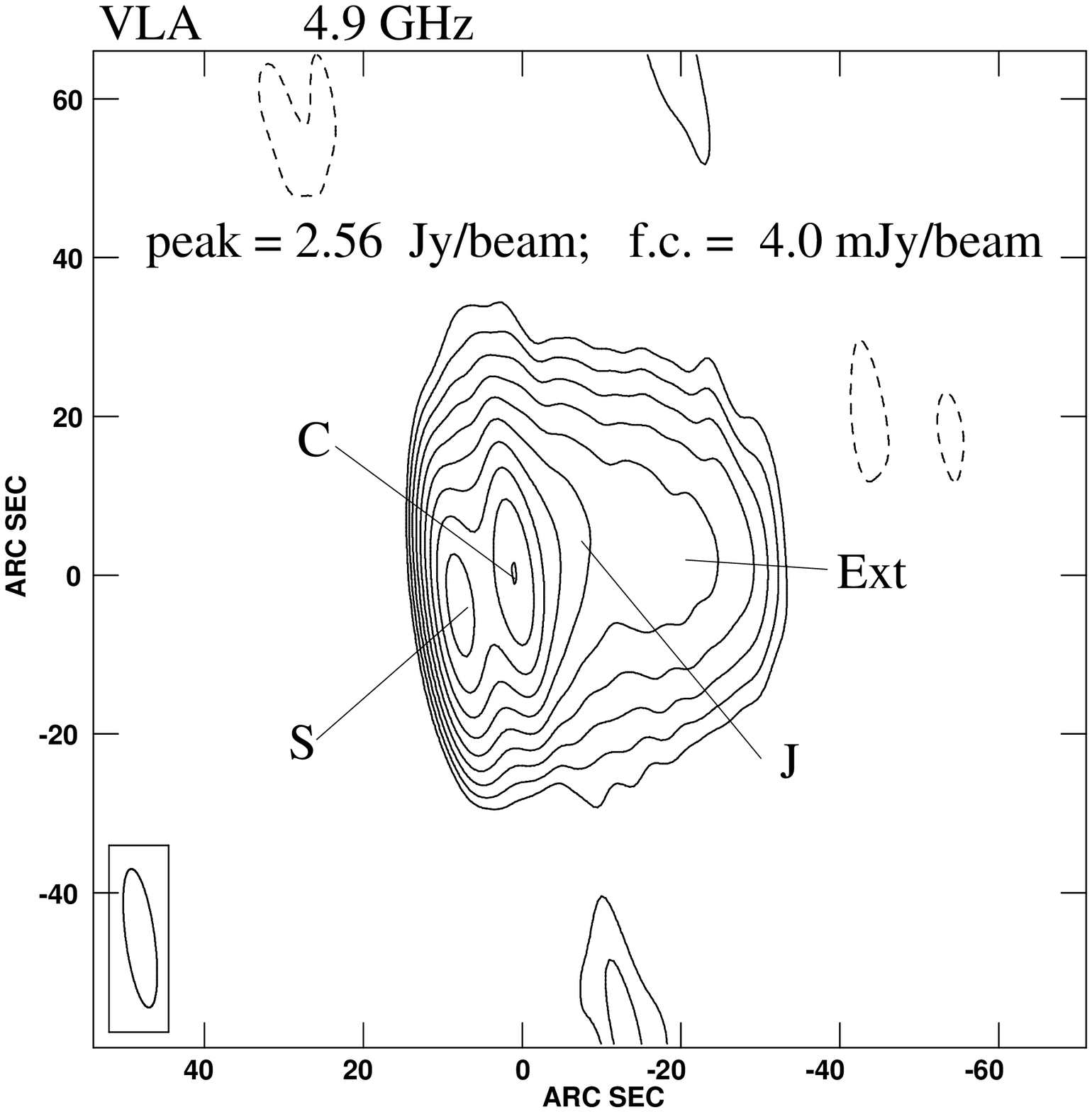}
\caption{VLA image of PKS 0521$-$36 at 4.9 GHz. On the image we provide the restoring beam, plotted in the bottom-left corner, the
peak flux density in Jy/beam, and the first contour (f.c.) intensity in
mJy/beam, which is three times the off-source noise level. Contour levels
increase by a factor of 2.}\label{VLA5}
\end{figure}

\begin{figure}
\includegraphics[width=7.5cm]{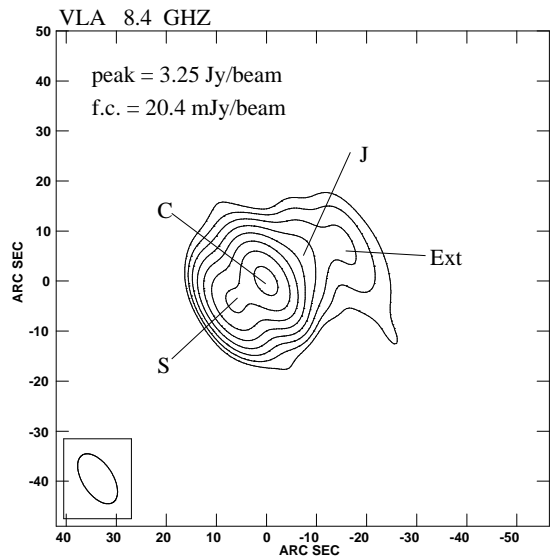}
\caption{VLA image of PKS 0521$-$36 at 8.4 GHz. On the image we provide the restoring beam, plotted in the bottom-left corner, the
peak flux density in Jy/beam, and the first contour (f.c.) intensity in mJy/beam, which is three times the off-source noise level. Contour levels
increase by a factor of 2.}\label{VLA8}
\end{figure}

In Figs.~\ref{VLA5} and \ref{VLA8} we show the total intensity images at 4.9
and 8.4 GHz, respectively. The radio emission is dominated by two compact
components, labelled C and S,
whose flux density is about 80 per cent and 73 per cent of the source total flux density at 4.9 and 8.4
GHz, respectively. The central component, coincident with the optical
nucleus, has an inverted spectrum ($\alpha_{4.9}^{8.4}$ $\sim$ --0.1$\pm$0.1) and hosts the
source core. This value is larger than that derived from VLBA data and this is
likely due to the contribution of the jet that cannot be separated from the
core component due to the lower resolution of the VLA observations. A jet-like structure, labelled J, emerges from 
the core with a position angle of about $-$45$^{\circ}$, in agreement
with what is found in high-resolution VLBA images (Fig.~\ref{vlba15}). At
about 5 arcsec (5 kpc) from the core, the jet bends toward the west, producing an extended structure of about
30 arcsec ($\sim30$~kpc) in size, in agreement with what is found at
lower frequencies \citep{falomo09, ekers89}. 
Component S is located about 7 arcsec ($\sim7$~kpc)
from the central component with a position angle of
120$^{\circ}$ and it marks the location of the hotspot. Its spectral
index is rather steep $\alpha_{4.9}^{\rm 8.4}$ = 1.0$\pm$0.1 and it is likely due
to the low resolution of these observations, which does not allow us to
disentangle the contribution of the hotspot from that of the lobe. It
is worth noting that the spectral indices have been computed assuming
the total flux density measured on images with different
resolution, which may produce an artificial steepening of the
spectrum. The errors on the spectral index have been computed by means
of the error propagation theory.

\begin{table}
\caption{VLA flux density and spectral index of PKS\,0521$-$36.}
\begin{center}
\begin{tabular}{cccc}
\hline
Comp & $S_{\rm 4.9}$ & $S_{\rm 8.4}$ &$\alpha_{\rm 4.9}^{8.4}$ \\
\hline
     &  Jy         & Jy          &                \\
\hline
C    & 3.72$\pm$0.18&  3.94$\pm$0.20& --0.1$\pm$0.1\\ 
S    & 2.18$\pm$0.11&  1.35$\pm$0.07&   1.0$\pm$0.1 \\
Ext  & 1.12$\pm$0.06&  0.35$\pm$0.03&   2.1$\pm$0.2 \\
Tot  & 7.02$\pm$0.35&  5.64$\pm$0.28&   0.4$\pm$0.1 \\
\hline
\end{tabular}
\label{vla_table}
\end{center}
\end{table}

\subsection{UMRAO data}

UMRAO centimeter band total flux density observations were obtained with
the University of Michigan 26 m paraboloid located in Dexter, Michigan, USA.
The instrument is equipped with transistor-based radiometers operating at
frequencies centered at 4.8, 8.0, and 14.5 GHz with bandwidths of 0.68, 0.79,
and 1.68 GHz, respectively. Dual horn feed systems are used at 8 and 14.5 GHz. Each observation
consisted of a series of 8 to 16 individual measurements over
approximately a 25 to 45 min time period, utilizing an ON-ON technique (switching the
target source between the two feed horns which are closely spaced on the sky)
at 8.0 and 14.5 GHz. As part of the observing procedure, drift scans were made
across strong sources to verify the telescope pointing correction curves, and
observations of nearby calibrators were obtained every 1 to 2 hours to correct
for temporal changes in the antenna aperture efficiency. Further details about
UMRAO are reported in \citet{aller14}. UMRAO data at 8 GHz and 14.5 GHz are represented in Fig.~\ref{MWL}. 

\subsection{SMA data}

\begin{figure*}
\centering
\vspace{10pc}
\includegraphics[width=14.0cm]{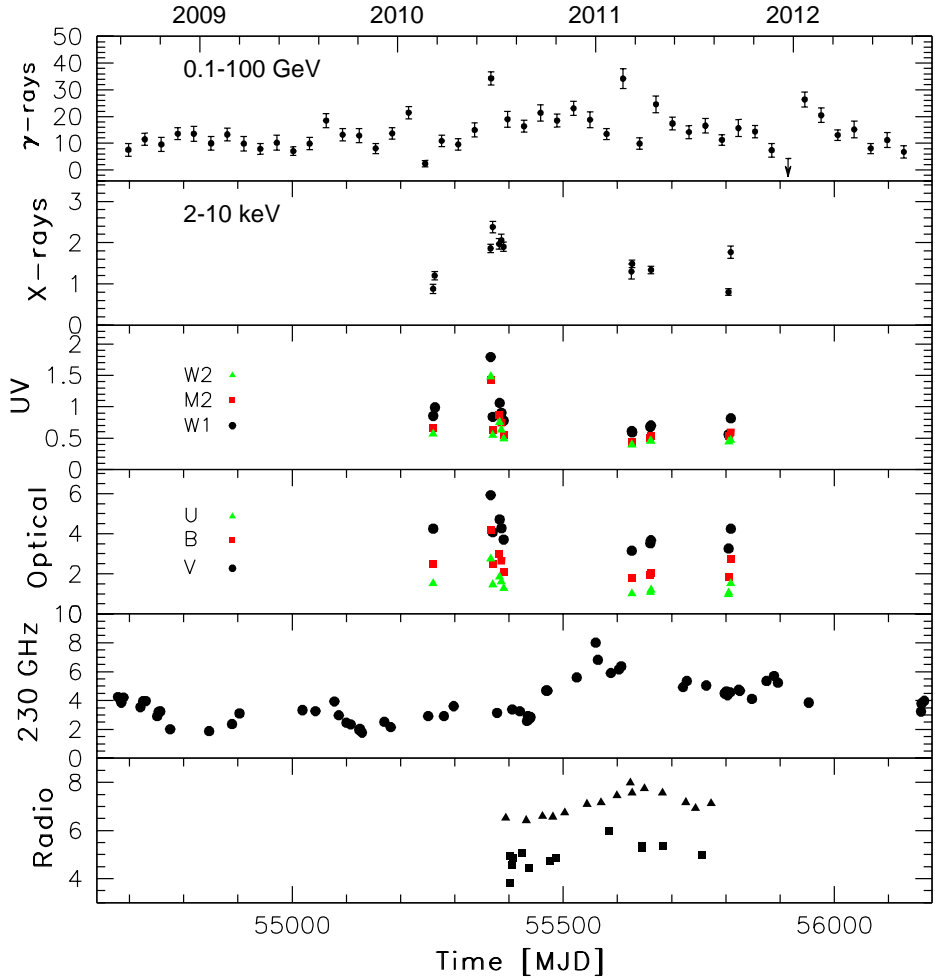}
\caption{Multifrequency light curve for PKS 0521$-$36. The period covered is 2008 August 4--2012 August 4 (MJD 54682--56143). The data were
  collected (from top to bottom) by {\em Fermi}-LAT ($\gamma$-rays; in units
  of 10$^{-8}$ ph cm$^{-2}$ s$^{-1}$), {\em Swift}-XRT (X-rays; in units
  of 10$^{-11}$ erg cm$^{-2}$ s$^{-1}$), {\em Swift}-UVOT ($w1$, $m2$, and
  $w2$ filters; in mJy), ($v$, $b$, and $u$ filters; in mJy), SMA (230 GHz; in
  Jy), UMRAO (8 and 15 GHz, triangles and squares, respectively; in Jy). The
  LAT data were obtained with the photon index left free to vary (see Fig.~\ref{Fig3}).}
\label{MWL}
\end{figure*}

The 230 GHz (1.3 mm) light curve was obtained at the SMA on Mauna Kea
(Hawaii). PKS 0521$-$36 is a bright AGN included in an ongoing monitoring program at the SMA to determine the fluxes of compact extragalactic radio sources that can be used as 
calibrators at mm wavelengths. Details of the observations and data reduction
can be found in \citet{gurwell07}. Data from this programme are updated
regularly and are available at the SMA website\footnote{http://sma1.sma.hawaii.edu/callist/callist.html. Use in
publication requires obtaining permission in advance.}. SMA data at 230 GHz
are shown in Fig.~\ref{MWL}.

\section{Discussion}

\subsection{Identification of the $\gamma$-ray source}\label{identification}

The association of a $\gamma$-ray source with a low-energy counterpart is
fundamental for understanding its physical properties.
The accuracy with which most of the $\gamma$-ray sources were localized with
EGRET was not enough to associate them with a known counterpart, leaving the
legacy of a large fraction of unidentified sources in $\gamma$-rays. Thanks to larger photon statistics and unprecedented angular
resolution at high energies {\em Fermi}-LAT is able to investigate the natures and counterparts
of these unidentified sources, although the $\gamma$-ray activity state as well as
correlated flux variability at different frequencies play an important
role for the identification process of a $\gamma$-ray source
\citep[e.g.,][]{dammando12}.

A $\gamma$-ray source at a position compatible with PKS 0521$-$36 was
detected with a 4.3$\sigma$ significance by EGRET during 1992 May 14--June 4,
with a flux of (1.8 $\pm$ 0.5)$\times$10$^{-7}$ ph cm$^{-2}$ s$^{-1}$ and a
photon index $\Gamma_{\gamma}$ = 2.16 $\pm$ 0.36 \citep{lin95}. As a result of
that activity the source was included in the Second EGRET catalogue as 2EG
J0524$-$3630 \citep{thompson95}. The $\gamma$-ray source 3EG J0530$-$3626 was included in
the Third EGRET catalogue, but PKS 0521$-$36 was outside the 99 per cent confidence contour in that catalogue, and the association between the $\gamma$-ray source and the AGN was not confirmed \citep{Hartman99}. The same conclusion was drawn for the $\gamma$-ray
source EGR J0529$-$3608 reported in the revised EGRET catalogue proposed by \citet{casandjian08}.

No $\gamma$-ray source associated with PKS 0521$-$36 was reported in the LAT Bright Source List obtained after the first
three months of {\em Fermi} operation \citep{abdo09}, but the AGN was associated with
1FGL J0522.8$-$3632, 2FGL J0523.0$-$3628, and 3FGL J0522.9$-$3628 in the
following {\em Fermi} catalogues \citep{abdo10a,nolan12,acero15}. 

Most of the associations of $\gamma$-ray sources with AGN in the {\em Fermi}
catalogues have high probability of being true, but a firm identification
needs to be confirmed by means of correlated variability in different energy bands.
The contemporaneous increase of activity observed in the optical, UV, X-ray, and
$\gamma$-ray bands in 2010 June (Fig.~\ref{MWL}) is a distinctive signature for the identification of the $\gamma$-ray source with PKS 0521$-$36. In the
following, we investigate the properties of PKS 0521$-$36 from radio to
$\gamma$-ray energy bands with the goal of unveiling the nature of this $\gamma$-ray emitting AGN.

\subsection{Gamma-ray properties}

By comparing the $\gamma$-ray properties of PKS 0521$-$36 with those of the
different types of AGN detected by LAT, we noted that the average photon index
of this source ($\Gamma_{\gamma}$ = 2.45 $\pm$ 0.02) is similar to the average
value observed for FSRQ \citep[$\Gamma_{\gamma}$ = 2.44 $\pm$
0.20;][]{ackermann15} and misaligned AGN \citep[$\Gamma_{\gamma}$ = 2.42 $\pm$
0.28;][]{Abdo2010}. 
As shown in Fig.~\ref{Fig3}, there is no general trend between $\gamma$-ray
flux and photon index for PKS 0521$-$36, but a significant hardening of the spectrum
was observed during the $\gamma$-ray flaring activity in
2010 June. This behaviour is in common with bright FSRQ and
low-synchrotron-peaked BL Lacs detected by LAT \citep{abdo10b}. The hardening
of the spectrum together with the low redshift make this source a promising
target for the current (i.e., MAGIC, VERITAS, and H.E.S.S.) and next generation of imaging atmospheric Cherenkov
telescopes (e.g., the Cherenkov Telescope Array). It is interesting to note that one of the few radio galaxies
detected at Very High Energy (VHE; E $>$ 100 GeV), NGC 1275 \citep{aleksic12},
showed flux and spectral index similar to those of PKS 0521$-$36 during the
2010 June flare. 

Contrary to the photon index, the average apparent isotropic $\gamma$-ray
luminosity of PKS 0521$-$36 (L$_{\gamma}$ = 5.2$\times$10$^{44}$ erg s$^{-1}$)
seems to lie in the region usually occupied by BL Lacs and radio galaxies
\citep{ackermann15}. However, the lower apparent luminosity of PKS 0521$-$36 with
respect to both FSRQ and the steep spectrum radio quasars (SSRQ) 3C 380 and 3C 207 may be due to the close
proximity of the source. In fact, a $\gamma$-ray flaring activity with peak
flux of 1--2$\times$10$^{-6}$ ph cm$^{-2}$ s$^{-1}$, as observed for PKS 0521$-$36,
is quite uncommon in BL Lacs \citep[e.g.,][]{cannon10} and in misaligned AGN, the only exception being the radio galaxy NGC 1275 \citep[e.g.,][]{ciprini13}.

It is not clear whether the most powerful jets of the
two parent populations (FSRQ and FR II radio galaxies) have analogous
structures. The difference in the jet structure may be related to the
different environment or to jet properties. We noted that the FR II radio galaxy Cygnus A, one of the most powerful radio sources with a redshift
comparable to that of PKS 0521$-$36, has not been detected in $\gamma$-rays so
far. On the other hand, Pictor A, an FR II radio galaxy at redshift $z$ =
0.035, was detected by {\em Fermi}-LAT \citep{acero15}, but with an apparent $\gamma$-ray
luminosity that is two order of magnitudes lower than that observed for PKS
0521$-$36. This may support the scenario of a complex jet structure in radio galaxies
(e.g., velocity gradients, spine-layer) with
possible differences in the emission mechanisms at higher energies also among
objects of the same class. Alternatively, the different viewing angle with
respect to the observer, and thus the different Doppler boosting of the jet
emission, may lead to the detection or not of radio galaxies in $\gamma$-rays.

No significant difference was observed between the average $\gamma$-ray flux detected by EGRET in 1992 and
LAT during 2008--2012. This is different from the long-term variability
observed for the radio-galaxy NGC 1275 between the EGRET and {\em Fermi}
era. In that case, the early LAT observations revealed a flux at least seven
times higher than the upper limit obtained by EGRET \citep{kataoka10}.
Of particular interest is the short variability observed
during flaring periods with a doubling time scale of the order of a few hours,
compatible with the variability observed in some bright FSRQ \citep[e.g.,][]{saito13,tavecchio10,abdo10c}. If the emitting region fills the entire cross
section of the jet, this rapid variability suggests a compact $\gamma$-ray
emitting region and the location of the $\gamma$-ray emission may be within
the broad-line region (BLR). Following the causality argument and considering the
12-hr variability observed during the $\gamma$-ray flares the size of the emitting region
should be $R < c t_{var} \delta /(1+z)$ = 6.3 $\times$10$^{15}$ cm (assuming
$\delta$ = 5; see Section~\ref{SEDmodeling}). This size is smaller than the values usually inferred from the SED
modeling of blazars within the framework of the standard one-zone leptonic model \citep[e.g.,][]{ghisellini11}. Alternatively, episodes of fast variability can
be produced in a jets-within-jet/mini-jets scenario \citep[e.g.][]{ghisellini08, giannios09} and in this case the location may be
further out, as proposed e.g. for M\,87 \citep[e.g.][]{Giroletti2012}, or
produced by turbulent cells in the relativistic plasma \citep{marscher14}.

\subsection{Radio properties}

On a kpc scale, the radio source shows a two-sided structure where jets, lobes,
and a hot spot are clearly visible in addition to the core.
The structure revealed by the VLBA images agrees with those reported in
previous works \citep[e.g.][]{tingay02,giroletti04}. However, since the data
were taken at distant epochs and at different frequencies, it is difficult to
make a reliable identification of the individual components. The best possible
match is probably on the bright diffuse region at $\sim30$ mas, whose flux
density and position are consistent in the various datasets, and essentially
stationary within the uncertainty related to its large size ($\sim
8$~mas). This bright resolved structure is  reminiscent of two other
$\gamma$-ray emitting AGN somewhat different from the typical blazars, i.e.\
M\,87 \citep{Abdo2009} and 3C 120 \citep{Abdo2010,Kataoka2011}. On parsec scales,
both sources show a one-sided jet whose brightness profile along the axis
initially decreases down to the noise level and then steeply rises in an
extended, stationary feature. The projected distance of this feature is
$\sim50$~pc in 3C 120 \citep{Roca-Sogorb2010} and $\sim70$~pc in M\,87
\citep{Giroletti2012,cheung07}. In PKS\,0521$-$36, a bright, extended jet
feature is found at nearly 30~pc (projected). These features may be
interpreted as recollimation shocks, and at least in the case of M87 it has
been proposed that high-energy emission up to the TeV regime might originate
at this location \citep{bromberg09}. Recently, a similar scenario was suggested
also for the rapid VHE flaring activity of the FSRQ 4C 21.35
\citep{tavecchio11}, although different models are proposed \citep{ackermann14,dermer12, tavecchio12}.

We estimated the ranges of viewing angles $\theta$ and jet velocity $\beta$
($=v/c$) from the jet/counter-jet brightness ratio from the radio images
presented in Sec.~\ref{vlbi}. Assuming the off-nuclear peak on the counter-jet
side as an upper limit to any possible counter-jet brightness, we calculate a
jet-counterjet brightness ratio $R\sim130$. In the standard hypothesis that
jet and counter-jet are intrinsically similar and that any asymmetry is
entirely due to relativistic beaming, this places limits on the jet velocity
and orientation by

\begin{equation}
R = \frac{B_{\rm j}}{B_{\rm cj}} = (\frac{1+\beta\,cos\theta}{1-\beta\,cos\theta})^{2+\alpha}
\end{equation}

\noindent and we obtain $\beta > 0.75$, $\delta > 1.5$, $\theta < 41^\circ$, assuming $\alpha=0.7$.

The radio spectrum below 1 GHz is dominated by the flux density
arising from the extended structures. The spectral index computed
between 160 MHz \citep{slee77} and 1.4 GHz from the NRAO VLA Sky Survey
\citep{condon98}
results to be 0.7. This radio spectral shape and the optical emission
suggest that PKS 0521$-$36 may be an intermediate object between a broad-line radio galaxy (BLRG) and an SSRQ.\\
We computed the core dominance (CD) of PKS\,0521$-$36 following the
definition in \citet{orr82,abdo10b}:

\begin{equation}
CD = \frac{S_{\rm core}}{S_{\rm tot} - S_{\rm core}} .
\end{equation}

\noindent where the core flux density $S_{\rm core}$ and the total
flux density $S_{\rm tot}$ refer to the source rest frame ($S = S_{\rm
obs}\times(1+z)^{\alpha_{\rm r} - 1}$).
For the core flux density we consider the value measured on the VLA
image at 4.9 GHz and reported in Table 6, and we assume a flat spectrum
$\alpha = 0$. In order to prevent missing
flux density associated with extended emission, we consider the total
flux density measured by the Parkes telescope at 5 GHz, which is
$S_{\rm tot}$ = 8.1 Jy \citep{wright96}, and we assume a spectral index
$\alpha=0.7$. We obtain log(CD) = $-$0.10. This value is similar to those
shown by the $\gamma$-ray emitting SSRQ 3C\,380 and 3C\,207 and BLRG 3C 111 \citep{abdo10b}.

In order to estimate the radio power and to make a comparison with the
$\gamma$-ray emitting misaligned AGN, we computed the radio luminosity at 178 MHz by using:

\begin{equation}
L = 4\pi d_{\rm L}^{2} S (1+z)^{\alpha_{\rm r} -1}
\end{equation}

\noindent where $d_{\rm L}$ is the luminosity distance, and S is the flux density at 178 MHz, which has been extrapolated from the values reported in the literature and corresponds to 67.5 Jy (source rest frame). The corresponding luminosity is L = 4.8$\times$10$^{26}$ W/Hz, which is a few orders of magnitude lower than the luminosity of the SSRQ 3C 380 and 3C 207, but is comparable with the luminosity of the BLRG 3C 111.

\subsection{X-ray to mm behaviour}

During 2005--2011 the source was monitored by the {\em Swift} satellite only sporadically, but a quite prominent flare was observed from optical to X-rays in 2010 June, contemporaneous with the peak of $\gamma$-ray flaring activity (see
Section \ref{identification}). The small variability amplitude (calculated as
the ratio of maximum to minimum flux) observed in X-rays ($\sim$ 3) with respect to that
observed in $\gamma$-rays ($\sim$ 13 on a monthly time-scale up to a factor of
$\sim$ 50 on a daily time-scale) together with the lack of hardening of the
spectrum may be an indication that the X-ray
emission is produced by the low-energy tail of the same electron distribution,
as observed in FSRQ \citep[e.g., PKS 1510$-$089;][]{dammando11}. On a time-scale of years a change by a factor of $\sim$10 in flux was detected between the {\em Chandra} and {\em Swift} observations. A second less prominent increase of
the X-ray activity was
observed on 2011 September 5, with no evident counterpart in
$\gamma$-rays. This may be due to different mechanisms or emission sites that give significant
contribution in X-rays and $\gamma$-rays in that period. 
The photon indices observed by {\em Swift}-BAT and {\em Swift}-XRT in hard
X-rays ($\Gamma_{\rm\,X}$ $\sim$1.8) and
soft X-rays ($\Gamma_{\rm\,X}$ $\sim$1.7) indicates that the peak of the IC emission should be at
MeV energies, according to the spectra observed by {\em Fermi}-LAT ($\Gamma_{\gamma}$ $>$ 2).

The variability amplitude in the optical-UV bands covered by {\em Swift}-UVOT
is $v$ $\sim$1.9, $b$ $\sim$2.4, $u$ $\sim$2.9, $w1$ $\sim$3.2, $m2$
$\sim$3.2, and $w2$ $\sim$3.7. We observed a larger variability amplitude at higher frequencies, contrary to what is observed in quasars,
where the contribution from the thermal accretion disc significantly
increases in the UV part of the spectrum, diluting the jet emission
\citep[e.g.][]{raiteri12}. This suggests the lack of a prominent accretion
disc in the broad-band spectrum of the source. We noted that during the
flaring activity in 2010 June the optical and UV emission seems to peak a few
days before the X-ray peak. This may be due the coverage provided by the {\em Swift} observations, and thus the lack of an observation at the time of the daily $\gamma$-ray peak, or to the complex jet structure with different parts of the jet
responsible for the emission at different frequencies.

The SMA data at 230 GHz show an amplitude variability of $\sim$4, with a peak flux density of 8
Jy on 2010 December 30 (MJD 55560).
This event at millimeter wavelengths may be the delayed flare with respect to the $\gamma$-ray flare observed in 2010
June. 
The increase of flux density was observed also by UMRAO at 15 and 8 GHz, with
the difference in time of about 1 month and 2 months, respectively, with respect to
the 230 GHz emission likely related to opacity effects due to
synchrotron self-absorption. Recently, \citet{fuhrmann14} investigated the
correlation between radio and $\gamma$-ray emission for a sample of bright
{\em Fermi} blazars. Performing a stacking analysis of the objects in the
sample, they found at 142 GHz a time-lag of (7$\pm$9) days between the radio and $\gamma$-ray
emission, with the $\gamma$-ray leading the radio emission. In this context a delay of six months between the peak at 230 GHz and the $\gamma$-ray
one for PKS 0521$-$36 is extremely large and thus unlikely, opening alternative
scenarios. The emitting region responsible for the optical-to-$\gamma$-ray
emission may be different from the region producing the mm and radio
emission. However, we note that SMA data are not available in 2010 May--June,
and therefore we cannot rule out a mm flare in the same period of the $\gamma$-ray flare.

\subsection{SED modeling}\label{SEDmodeling}

We build two SED for PKS 0521$-$36 in two different activity states: the high
state in 2010 June and a low state in 2010 February-March. The flaring SED
includes the {\em Fermi}-LAT spectrum built with data centred on 2010 June 22--July 5 (MJD
55369--55382), the optical, UV, and X-ray data collected by {\em Swift} on
2010 June 23 (MJD 55370), and 230 GHz data collected by SMA on 2010 July 1 (MJD 55378). The low SED includes the
LAT spectrum built with data centred on 2010 February 1--March 31 (MJD 55228--55286), the optical, UV, and X-ray data collected by {\em Swift} on
2010 March 5 (MJD 55260), and 230 GHz data collected by SMA on 2010 February 24 (MJD 55251). 

The modeling of the SED of PKS 0521$-$36 within the framework of the standard
one-zone leptonic model provides results at odds with the bulk of the blazar
population. In particular, a quite low Doppler factor is required
\citep[$\delta$ $\simeq$ 3; e.g.,][]{pian96, ghisellini11}, implying a
relatively large viewing angle $\theta _{\rm v}\gtrsim 15$$^{\circ}$ and a
low bulk Lorentz factor. This result can be traced back to the relatively
small separation between the two bumps in the SED. This conclusion is
corroborated by several hints, such as the existence of a large-scale optical jet and the
moderate core-to-extended radio flux ratio. In the case of a relatively large
viewing angle and thus small beaming amplification, it is not excluded that
also slow regions of the jets contribute to the observed emission. This would
be the case in the framework of the {\it structured jet} scenario, in which
the flow is supposed to be composed of a fast inner core -- or spine --
surrounded by slower plasma -- the layer \citep{ghisellini05,tavecchio08}. Indeed such a scheme has been applied to interpret the
emission of $\gamma$-ray emitting radio galaxies, for which the emission can
be interpreted as a mix of spine and layer components \citep[see also][]{tavecchio14}.
With this motivation, in the following we focus our attention on the application of the structured jet model to the emission of PKS 0521$-$36. 

We briefly recall the main features and the parameters of the model (refer to Ghisellini et al. 2005, Tavecchio \& Ghisellini 2008, 2014 for a more complete description). The spine is assumed to be a cylinder with height $H_s$ and radius $R_s$ (the subscripts `s' and `l' stand for spine and layer, respectively). The layer is modeled as a hollow cylinder  with height  $H_l$, internal radius $R_s$ and 
external radius $R_l=1.2\times R_s$.  We assume $H_s\ll H_l$, ideally corresponding to the case of a perturbation traveling down the spine, surrounded by a relatively long and stationary layer of slow plasma.

% --------------------------------------------------------
\begin{figure}
%\hskip -2 cm
\hspace{-3.5 truecm}
%\vspace*{-2.5 truecm}
%\vskip -0.3 cm
\psfig{file=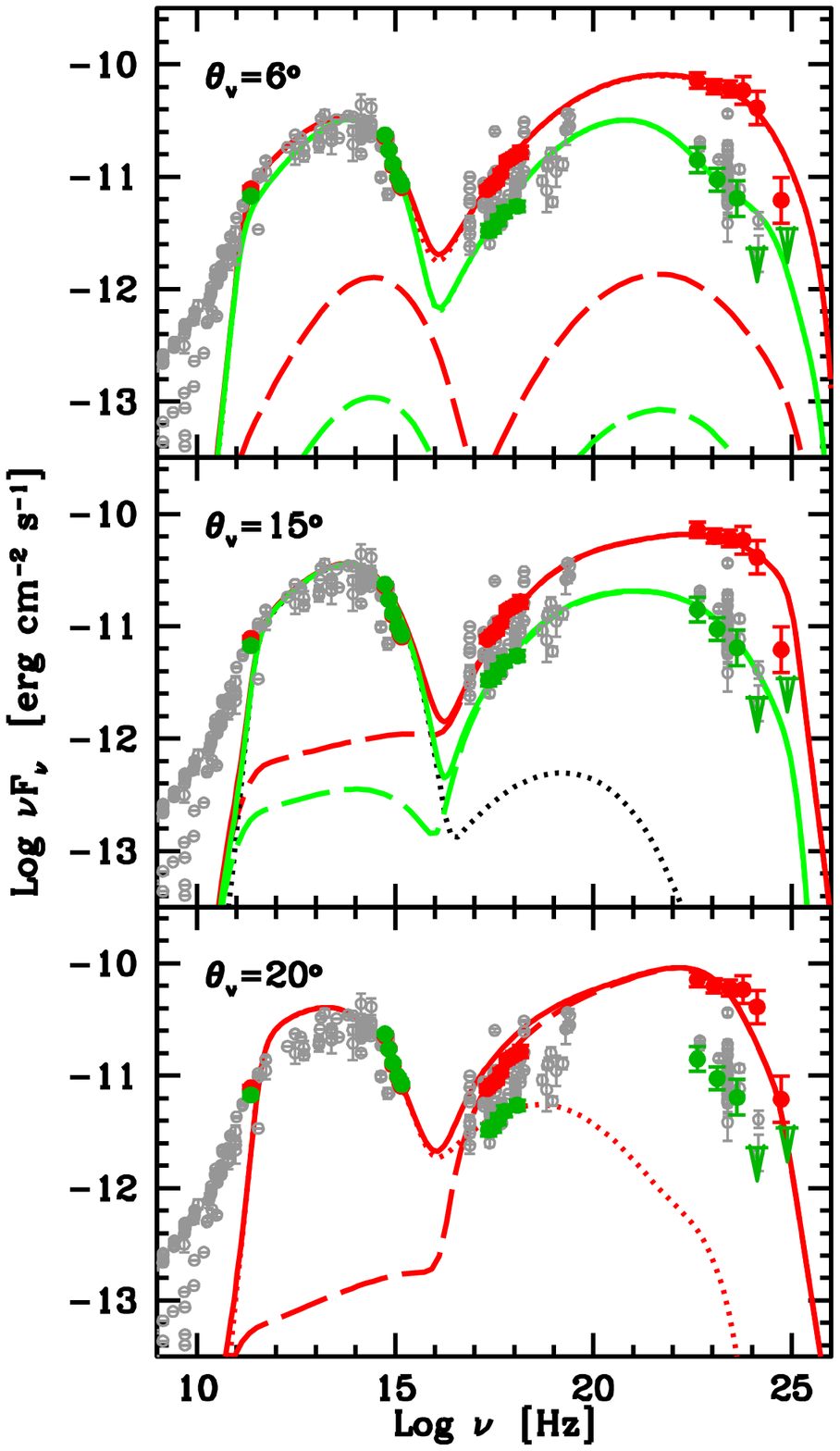,height=16.cm,width=16.cm}
%\vspace{-4.4 cm}
\caption{
Spectral energy distribution of PKS 0521$-$36. 
Red symbols show the high state, green symbols the low state. Arrows
represent 2$\sigma$ upper limits. Historical data (gray) are from the ASDC archive ({\tt http://tools.asdc.asi.it}).
The three panels report the results obtained with the structured jet emission model for different values 
of the viewing angle, from top to bottom: 
$\theta_{\rm v}=6^{\rm o}$, $\theta_{\rm v}=15^{\rm o}$ and $\theta_{\rm v}=20^{\rm o}$. 
The dashed and the dotted lines show the emission from the layer and the spine, respectively. 
The thick solid line shows the total. For the case $\theta_{\rm v}=15^{\rm o}$ the same spine emission (shown in black)
is used for both states. See text for details.
}
\label{sed}
\end{figure}
% --------------------------------------------------------

% -----------------------------------------------------------
\begin{table*} 
\begin{center}
\begin{tabular}{ll|llllllllllll|}
\hline
&$R$    
& $H$  
& $L_{\rm syn}$  
& $B$  
& $\gamma_{\rm min} $  
& $\gamma_{\rm b} $ 
& $\gamma_{\rm max}$  
& $n_1$
& $n_2$
& $\Gamma$ 
& $\theta_{\rm v}$ \\
& cm  &cm &erg s$^{-1}$ & G & & &  & & & &deg \\
\hline  
Layer low &$3.5\times 10^{16}$ &$2\times 10^{16}$ &$6\times10^{39}$  &0.5 &100 &$3\times10^3$ &$7\times10^4$   &1.5   &3   &3 &6 \\ 
Spine low & $3.5\times 10^{16}$ &$10^{15}$ &$1.9\times10^{41}$  &0.18 &60 &$1.2\times 10^4$ &$2.55\times10^4$ &2 &3.7 &12  &6  \\
Layer high &$3.5\times 10^{16}$ &$2\times 10^{16}$ &$7\times10^{40}$  &0.5 &100 &$3\times10^3$ &$7\times10^4$   &1.5   &3   &3 &6 \\ 
Spine high & $3.5\times 10^{16}$ &$10^{15}$ &$2.2\times10^{41}$  &0.17 &60 &$6\times 10^3$ &$3\times 10^4$ &2.1 &3 &12  &6  \\
\hline  
Layer low & $5\times 10^{16}$ &$2\times 10^{16}$ &$1.3\times10^{41}$  &1 &80 & $3\times10^4$&$10^5$   &2.75   &4   &3 &15 \\ 
Layer high & $5\times 10^{16}$ &$2\times 10^{16}$ &$5.0\times10^{41}$  &1 &100 & $10^4$ &$5\times10^5$   &2.7   &4  &3 &15 \\ 
Spine$^*$ & $5\times 10^{16}$ &$10^{15}$ &$4\times10^{43}$  &15 &10 &$2\times 10^3$ &$6\times 10^3$ &2.1 &3 &10  &15  \\
\hline
Layer high & $5\times 10^{16}$ &$2\times 10^{16}$ &$2\times10^{41}$  &1 &80 & $6\times10^4$ &$8\times10^6$   &2.7   &4   &3 &20 \\ 
Spine high & $5\times 10^{16}$ &$10^{15}$ &$4\times10^{44}$  &15 &10 &$300$ &$8\times 10^3$ &2.1 &3 &10  &20  \\
\hline
\end{tabular}                                                         
\caption{Input parameters of the models for the layer and the spine
shown in Fig. \ref{sed}. All quantities (except the bulk Lorentz
factors $\Gamma$ and the viewing angle $\theta_{\rm v}$) are measured
in the rest frame of the emitting plasma.  The external radius of the
layer is fixed to the value $R_2=1.2 \times R$. $^*$For the case $\theta_{\rm v}=15^{\circ}$ the same spine emission has been used for both states.}
\label{tabparam}
\end{center}
\end{table*}                                                                  
% --------------------------------------------------------------

In both zones we assume a phenomenological and stationary electron energy distribution described as a broken power law with the following parameters: minimum, maximum and break Lorentz factors $\gamma _{\rm min}$, $\gamma _{\rm max}$ and $\gamma _{\rm b}$ and indices $n_1$ and $n_2$. 
The normalization of the electron distribution is parametrized by the emitted synchrotron luminosity, $L_{\rm syn}$. The emitting regions contain a tangled magnetic field $B_{s}$, $B_{l}$. 
The relativistic beaming is specified by the two Lorentz factors $\Gamma _{s}$, $\Gamma_{l}$ and by the viewing angle $\theta_{\rm v}$.

Electrons emit through synchrotron and IC mechanisms. Following \citet{ghisellini09}, $R_{\rm\,BLR}$ = 10$^{17}$$\times$$L_{\rm
  disc, 45}^{0.5}$ cm and for $L_{\rm\,disc}$ = 3$\times$10$^{43}$ erg s$^{-1}$ we
obtain $R_{\rm\,BLR}$ = 2$\times$10$^{16}$ cm. Therefore the emitting region
should be beyond the BLR and for this reason the external Compton of the BLR seed
photons should be negligible. For the IC besides the
local synchrotron radiation field (synchrotron self-Compton, SSC) we consider the beamed radiation field of the other component. 
Indeed, because of the relative motion, the energy density of radiation produced in one component is 
boosted by the square of the relative Lorentz factor $\Gamma_{\rm rel}$ in the 
rest frame of the other, where $\Gamma_{\rm rel}=\Gamma_{\rm s} \Gamma _{\rm l}(1-\beta_{\rm s}\beta_{\rm l})$. As we will see, absorption of $\gamma$-rays through the interaction with the soft radiation field, $\gamma \gamma\rightarrow e^+ e^-$ can be important.
In the case under study, radiation is produced and absorbed within the same
region and thus the ``suppression factor" is $I(\nu)/I_{\rm
  o}(\nu)=(1-\exp[-\tau_{\gamma \gamma}(\nu)])/\tau_{\gamma \gamma}(\nu)$
which, for large optical depths, is well approximated by $1/\tau_{\gamma
  \gamma}$. 

We note that the number of the parameters is large (almost twice that of the simple SSC model). However,
the model has to satisfy constraints that can be used as guidelines in
selecting the suitable setup. We refer to \citet{tavecchio14} for a detailed discussion.

Following the indications recalled in the previous subsections we consider
angles larger than those usually associated with blazar jets, $\theta_{\rm
  v}\lesssim 5^{\circ}$. In Fig.~\ref{sed} we show the results of the modeling of the SED with three different angles, $\theta_{\rm v}=6^{\circ}$, $15^{\circ}$ and $20^{\circ}$. The corresponding parameters are reported in Table~\ref{tabparam}.

We expect that, for relatively small angles, the spine radiation is much more boosted than that of the layer, implying that the spine emission component largely dominates the observed SED. This solution is that assumed for the case 
$\theta_{\rm v}=6^{\circ}$ (upper panel). The relative stability of the
low-energy component accompanied by the variation which occurred in the IC
peak is obtained by changing the layer luminosity (and maintaining almost the
same parameters for the spine). In fact, the spine IC bump is dominated in the
high state by the component resulting from the IC scattering of the electrons in the spine with the photons coming from the layer. Variations of the layer luminosity are thus accompanied by substantial variations of the IC luminosity of the spine.

\begin{figure}
\hspace{-.3 truecm}
\psfig{file=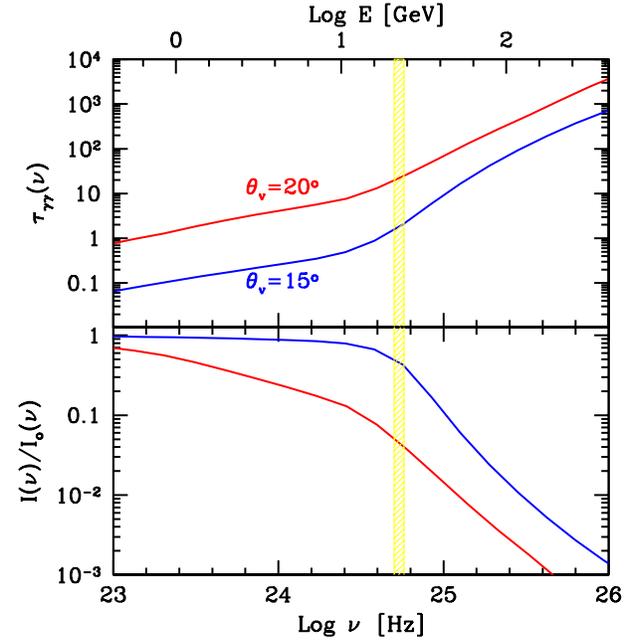,height=9.cm,width=9.cm}
\vspace{-.7 truecm}
\caption{Optical depth (upper panel) and suppression factor (lower panel) 
for absorption of $\gamma$-rays within the jet as a function of the frequency
for the models of the high states reported in Fig. \ref{sed}, for $\theta_{\rm v}=15^{\rm o}$ (blue) and $\theta_{\rm v}=20^{\rm o}$ (red). The vertical
yellow line shows the upper energy limit of the LAT spectrum.
}
\label{tau}
\end{figure}

For larger angles the only suitable solution is that the two bumps are dominated by the two jet components, i.e. the low energy bump by the synchrotron component of the spine and the high energy bump by the IC emission of the layer \citep[see the discussion in][]{tavecchio14}. For $\theta_{\rm v}=15^{\circ}$ (middle panel) 
the model still reproduces quite well the data. In this case we also assume a stationary spine and a variable layer component. A further increase of the angle (lower panel) is limited by a problem already discussed in \citet{tavecchio14}. Indeed, to compensate the reduced relativistic amplification of the spine emission one has to increase its intrinsic luminosity. This radiation field -- concentrated in the optical-IR band -- is the ideal target for the pair producing reactions $\gamma \gamma \to e^+ + e^-$ with $\gamma$-ray photons produced in the layer. 
In Fig.~\ref{tau}, we  show the optical depth $\tau_{\gamma \gamma}$ and the
corresponding ``suppression factor" $[1-\exp(-\tau_{\gamma \gamma})]/\tau_{\gamma \gamma}$ as a function of the frequency for the case
$\theta_{\rm v}=15^{\circ}$ and $\theta_{\rm v}=20^{\circ}$. The shapes of the
curves, related to the spine target photon spectrum, are similar. However, due to the larger intrinsic luminosity, for  $\theta_{\rm v}=20^{\rm o}$ the optical depth is larger by a factor of $\approx10$. As a consequence, while
$\theta_{\rm v}=15^{\rm o}$ the source is transparent ($\tau_{\gamma \gamma}<1$ up to $\sim 20$ GeV, corresponding to the highest-energy LAT bin in the high state, energy marked by the vertical yellow line), for $\theta_{\rm v}=20^{\rm o}$, the optical depth reaches unity already at 
$\approx 1$ GeV and rapidly increases, determining the abrupt cut--off visible
in Fig.~\ref{sed} (lower panel). We therefore conclude that, in the framework of the structured jet model, the angle cannot be larger than about $15^{\circ}$.

\begin{figure}
\hspace{-.5 truecm}
\psfig{file=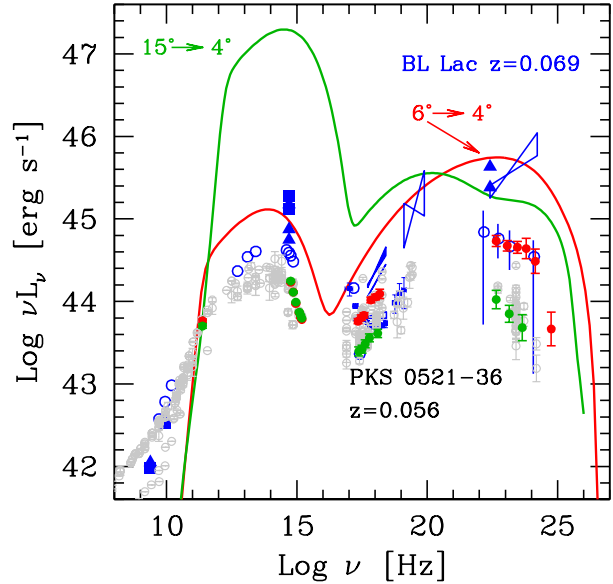,height=9cm,width=9.cm}
\vspace{-.7 truecm}
\caption{The green and the red lines show the emission from the spine-layer model of PKS 0521$-$36 for the cases $\theta_{\rm v}=15^{\circ}$ and  $\theta_{\rm v}=6^{\circ}$ as
observed at $\theta_{\rm v}=4^{\circ}$.  Red symbols show the high state,
green symbols the low state. LAT upper limits are not shown. For comparison, the blue data points describe the SED of  BL Lac itself.
}
\label{aligned}
\end{figure}

In the unification scheme for radio galaxies and blazars we expect that, once
observed at small angles, the SED of a radio galaxy resembles that of a
typical blazar. Therefore for the remaining two cases, $\theta_{\rm v}=6^{\circ}$ and $15^{\circ}$ one
can further ask what kind of SED would the source present if observed at
angles more typical for blazars. This is shown in Fig.~\ref{aligned}, in which
we report the SED as recorded by an observer at $\theta_{\rm v}=4^{\circ}$,
not changing the other parameters. For comparison we also display the
historical SED of BL Lacertae. The case $\theta_{\rm v}=15^{\circ}$ (green) results in an SED
strongly unbalanced (by a factor $\approx 100$) toward the synchrotron
luminosity, at odds with the typical blazar SED. This effect is clearly
related to the fact that the high-energy bump in the case  $\theta_{\rm v}=15^{\circ}$ is associated with the layer. At small angles, the layer
radiation is much less beamed than that of the spine and this determines the
substantial prominence of the low-energy peak. On the contrary, the SED for
case $\theta_{\rm v}=6^{\circ}$ (red) resembles that of the prototypical BL
Lac. This suggests that a viewing angle between $6^{\circ}$ and $15^{\circ}$ is
favored. 

\noindent In addition in the spine-layer scenario proposed here, the fast variability observed in $\gamma$-rays is
difficult to reconcile with the cases that have $\theta_{\rm v}=15^{\circ}$, in which the region responsible for the $\gamma$-ray
emission is relatively large (5$\times$10$^{16}$ cm) and the Doppler factor
small ($\delta$ = 4). Reducing the size of the emitting region in these two
cases should lead to an increase of the compactness of the region and thus of
the corresponding $\gamma$--ray opacity. This is another indication in favour
of a relatively small viewing angle.

However, we cannot rule out the possibility that PKS 0521$-$36 is an AGN seen
at $\theta_{\rm v}$ $\sim$15$^{\circ}$ and its `aligned' counterpart is a blazar
belonging to a population of synchrotron-dominated objects.

 % --------------------------------------------------------
% --------------------------------------------------------

% --------------------------------------------------------

\section{Conclusions}

We investigated the properties of PKS 0521$-$36, an AGN with uncertain
classification. We report results on multiwavelength observations of this
source obtained from radio through $\gamma$ rays by SMA, UMRAO, VLA, {\em Swift}
and {\em Fermi}-LAT, mostly during 2008 August--2012 August. In addition
archival {\em XMM--Newton}, {\em Chandra}, and VLBA observations were
analyzed. 

The historical classification of this source as a BL Lac object is
in contrast with the broad emission lines observed in optical and UV, with EW
larger than 5 \AA\, (rest-frame). The
presence of broad emission lines may suggest a classification as an FSRQ, but the radio
spectrum below 1.4 GHz is not flat ($\alpha_{\rm\,r}\sim$ 0.7). The radio
spectral shape and the optical emission lines indicate a possible classification as an intermediate object between BLRG and SSRQ. On pc-scales PKS 0521$-$36
shows a one-sided jet with a brightness distribution along the axis decreasing down
to the noise level and then rising in an extended, stationary feature at
nearly 30 pc. This is similar to what was observed in M\,87 and 3C 120, two $\gamma$-ray emitting misaligned AGN.
The core dominance estimated for PKS 0521$-$36 is similar to those of the
$\gamma$-ray emitting SSRQ 3C 380 and 3C 207 and BLRG 3C 111, suggesting an intermediate viewing angle with respect to the observer. 

The fit of the {\em XMM--Newton} EPIC pn spectrum confirmed the presence of a
contribution of the thermal component identified by \citet{birkinshaw02} in
the {\em Chandra} image as a diffuse halo around the AGN. No significant Fe K$\alpha$ emission was detected
in the {\em XMM--Newton} spectrum. This seems to disagree with the
possible classification of PKS 0521$-$36 as a BLRG, although a smaller viewing
angle with respect to the other BLRG may lead to a significant increase of the
jet component that overwhelms the Fe K$\alpha$ line in this object
\citep[similarly to the case of the narrow-line Seyfert 1 galaxy PMN J0948$+$0022;][]{dammando14}.

The contemporaneous flux increase observed from optical to $\gamma$ rays during 2010 June
suggests the identification of the $\gamma$-ray source with PKS
0521$-$36. The 230 GHz light curve showed an increase of activity peaking on
2011 January. A delay of $\sim$6 months between the emission at mm and
$\gamma$ rays is unlikely, suggesting that different parts of the jet are
responsible for those activities. However, we cannot rule out that a mm peak contemporaneous to the $\gamma$-ray one was missed due to the lack of
 observations at 230 GHz in 2010 May--June. The average $\gamma$-ray photon index of the source ($\Gamma_{\gamma}$ = 2.45
$\pm$ 0.02) is similar to the average value observed for FSRQ as well as for
the FR II radio galaxies (in particular the SSRQ). Moreover, the average apparent isotropic luminosity of the source (L$_{\gamma}$ = 5.2 $\times$10$^{44}$ erg s$^{-1}$) is in agreement with the
values observed for BL Lacs and radio galaxies. However, the lower luminosity with respect to SSRQ may be due to the close proximity of the source. Thus, taking into account its low redshift, the $\gamma$-ray properties of PKS 0521$-$36 are compatible with those observed in
SSRQ and BLRG. In this context the strong flaring activity detected from
this source by {\em Fermi}-LAT starting from 2010 June, with daily peak flux
of 1--2$\times$10$^{-6}$ ph cm$^{-2}$ s$^{-1}$, is very intriguing. 

We discuss the $\gamma$-ray emission of this source in the framework of the `spine-layer' scenario, with a fast spine surrounded by a
slower layer, like in the case of the radio galaxy NGC 1275
\citep{tavecchio14}. We compare the SED of the flaring state (2010 June) with
that of a low activity state (2010 February-March). We present three alternative models,
corresponding to three different choices of the viewing angles $\theta _{\rm v}$ = 6$^{\circ}$, 15$^{\circ}$, and
20$^{\circ}$. For the case with $\theta _{\rm v}$ = 6$^{\circ}$ and
15$^{\circ}$ we obtain a good fit. According to the unification scheme for
radio galaxies and blazars, reporting the SED obtained with $\theta _{\rm v}$ = 15$^{\circ}$ as recorded by an
observer at $\theta _{\rm v}$ = 4$^{\circ}$, this SED would correspond to that
of the `aligned' counterpart of the radio galaxy. On the contrary, the resulting SED is strongly unbalanced toward the synchrotron luminosity (by a factor of $\approx 100$), at odds with the typical blazar
SED. This suggests that a viewing angle between 6$^{\circ}$
and 15$^{\circ}$ is preferred. In the spine-layer scenario proposed here the rapid variability observed during $\gamma$-ray
flares favours a relatively small angle. However, we cannot rule out the possibility that PKS 0521$-$36 is an AGN seen
at $\theta_{\rm v}$ $\sim$15$^{\circ}$ and its `aligned' counterpart is a blazar
belonging to a population of synchrotron-dominated objects.

\noindent This intermediate viewing angle is in agreement with the core
dominance and the other radio properties observed, and thus with the classification of
PKS 0521$-$36 as an SSRQ at low redshift or a BLRG. The strong $\gamma$-ray flaring activity observed by
{\em Fermi}-LAT from this source indicates that SSRQ and BLRG may have relativistic
jets as powerful as blazars. We noted that an increase of the $\gamma$-ray
activity was observed for the SSRQ 3C 380 \citep{torresi13}, although at
a flux level lower than that reached by PKS 0521$-$36.  

The multifrequency observations and SED modeling presented here give
important information about the characteristics of the AGN PKS 0521$-$36 and
its classification. Further radio-to-$\gamma$-ray observations will be
fundamental to investigate in even more detail the nature and the physical
processes occurring in  this source.

\section*{Acknowledgments}

The {\em Fermi} LAT Collaboration acknowledges generous ongoing support from a
number of agencies and institutes that have supported both the development and
the operation of the LAT as well as scientific data analysis. These include
the National Aeronautics and Space Administration and the Department of Energy
in the United States, the Commissariat \`a l'Energie Atomique and the Centre
National de la Recherche Scientifique / Institut National de Physique
Nucl\'eaire et de Physique des Particules in France, the Agenzia Spaziale
Italiana and the Istituto Nazionale di Fisica Nucleare in Italy, the Ministry
of Education, Culture, Sports, Science and Technology (MEXT), High Energy
Accelerator Research Organization (KEK) and Japan Aerospace Exploration Agency
(JAXA) in Japan, and the K.~A.~Wallenberg Foundation, the Swedish Research
Council and the Swedish National Space Board in Sweden. Additional support for
science analysis during the operations phase is gratefully acknowledged from
the Istituto Nazionale di Astrofisica in Italy and the Centre National
d'\'Etudes Spatiales in France.             

We thank the {\em Swift}  team for making these observations
possible, the duty scientists, and science planners. The Submillimeter Array is a joint project between the Smithsonian
Astrophysical Observatory and the Academia Sinica Institute of Astronomy
and Astrophysics and is funded by the Smithsonian Institution and the
Academia Sinica. This research has made use of data from the University of
Michigan Radio Astronomy Observatory which has been supported by the
University of Michigan and by a series of grants from the Nation al Science
Foundation, most recently AST-0607523. This research was supported in part by NASA
Fermi Guest Investigator awards NNX09AU16G, NNX10AP16G, NNX11AO13G, and
NNX13AP18G. Part of this work is based on archival data, software or online
service provided by ASI Science Data Center (ASDC). The National Radio
Astronomy Observatory is a facility of the National Science Foundation
operated under cooperative agreement by Associated Universities, Inc. We thank
S. Digel, D. J. Thompson, J. Perkins, D. Gasparrini, and the anonymous referee for useful comments and suggestions.

\end{document}